\documentclass[amsmath,amssymb,aps,pra,superscriptaddress,twocolumn]{revtex4-2}
\usepackage{tikz}
\usepackage{amsmath}
\usepackage{mathtools}
\usepackage{physics}
\usepackage{dsfont}
\usepackage{setspace}
\usepackage{xcolor}
\usepackage{graphicx}
\usepackage{svg}
\usepackage{subcaption}
\usepackage{overpic}
\usepackage{textgreek}
\usepackage{amssymb}
\usepackage{comment}
\usepackage{tabularx}
\usepackage[labelformat=simple]{subcaption}
\usepackage{float}
\usepackage{hyperref}
\usepackage{titlesec}
\usepackage[toc,page]{appendix}
\renewcommand{\raggedright}{\leftskip=0pt \rightskip=0pt plus 0fil}

\begin{document}

\title{Dynamics and spectra of open quantum systems coupled to anharmonic environments}

\author{Kian Damezin}%
\affiliation{SUPA, Institute of Photonics and Quantum Sciences, Heriot-Watt University, Edinburgh, EH14 4AS, UK}

\author{Brendon W.~Lovett}%
\affiliation{SUPA, School of Physics and Astronomy, University of St Andrews, St Andrews KY16 9SS, United Kingdom}

\author{Erik M.~Gauger}%
\affiliation{SUPA, Institute of Photonics and Quantum Sciences, Heriot-Watt University, Edinburgh, EH14 4AS, UK}

\author{Adam Burgess}%
\affiliation{SUPA, Institute of Photonics and Quantum Sciences, Heriot-Watt University, Edinburgh, EH14 4AS, UK}
\begin{abstract}

In open quantum systems, the environment is typically modelled as a collection of harmonic oscillators, whereas realistic environments often exhibit unique non-Markovian effects due to the anharmonicity of the environment. Here we present a framework for modelling anharmonic environments in open quantum systems numerically exactly, in which, via a matching criterion, we map the anharmonic environment onto an effective harmonic one. This results in an effective spectral density that is a collection of shifted bare spectra arising from the non-equidistant energy gaps of the anharmonic environment, along with a unique zero-frequency term that effectively acts as static disorder on the system. This is due to the non-zero variance of the diagonal of the bath coupling operator. In tandem, we develop a framework for investigating the effects of environments comprising a few damped anharmonic modes where the matching criterion fails to work due to the non-Gaussianity of the environment. For both continuous and discrete damped anharmonic environments, this work investigates their unique influence on the system's dynamics, including enhanced non-Markovianity compared to a harmonic approximation, novel effects in the absorption spectrum, and the potential for anharmonic environments to enhance energy transport.
 
\end{abstract}
\maketitle
\section*{Introduction}

Open quantum systems, where a quantum subsystem interacts with an external environment, form the backbone of modern theoretical frameworks across many domains in quantum physics, quantum optics, chemistry, and materials science ~\cite{InesDeVega_Daniel_Alonso_Dynmaics_OQS,breuer2002theory}. In many cases, the environment is modelled as a collection of non-interacting harmonic oscillators -- a simplification that allows for elegant mathematical treatment and the use of powerful tools like the Feynman-Vernon influence functional~\cite{feynman1963theory}, the Caldeira-Leggett model~\cite{caldiera1983path}, various master equation approaches~\cite{breuer2002theory, linblad1976generators} and numerical exact methods ~\cite{DAMPF,chin_exact_2010, strathearn_efficient_2018, jorgensen_exploiting_2019,lacroix2026tensornetworkmethodsnonperturbative}. These models have been instrumental in our understanding of dissipation, decoherence and thermalization, particularly in biological systems~\cite{NicolaMicroscopicSimulations,schroder_tensor_2019} and quantum optical and solid state systems~\cite{LacorixPhotonicStructure, chekhovich_nuclear_2013, Dubertrand_solid_state}.

However, in many realistic systems---particularly in complex biomolecular, chemical, and material environments---the harmonic approximation can break down. The environmental degrees of freedom, ranging from nuclear motions in large biomolecules to vibrational modes in soft condensed phases, often exhibit significant anharmonicity~\cite{wang2003multilayer, shi2009efficient, tiwari2013electronic,Wang2007}. Such anharmonic features can drastically alter the system’s dynamics, introducing additional memory effects, nonlinear couplings, and non-Gaussian noise characteristics that challenge the assumptions of standard open quantum system approaches ~\cite{Yeganeh2006,Evans2000,WangThoss2004,ThossWang2006,Lopez2011}.

In the context of physical chemistry -- particularly in ultrafast spectroscopy, light-harvesting complexes, and quantum control of molecular processes-- capturing the correct bath dynamics is crucial for interpreting and predicting phenomena such as energy transfer, vibrational relaxation, coherent oscillations, and absorption and emission spectra~\cite{mukamel1995nonlinear,scholes2003long,cho2008coherent, Roos, dewit_extracting_2015}. The inclusion of anharmonicity in the environments has thus emerged as a frontier~\cite{MaxInesDeVega, CarlosDeVega,Smith2019QuantumDissipative} in theoretical and computational modelling, requiring the development of new methodologies that go beyond the harmonic approximation.

Recent advances in this area include the development of numerically exact tools capable of exploring anharmonic environments such as ML-MCTDH~\cite{wang2003multilayer} and the Automatic Compression of Environments algorithm (ACE)~\cite{mortiz2022ACE,Ye2021TensorNetworkIF}. These approaches capture the non-Markovian and anharmonic effects of the environment by explicit representation of the environment. This enables an accurate description of the interplay between system and environment in regimes where anharmonicity plays a pivotal role, opening new avenues for understanding complex quantum phenomena in physical chemistry and materials science.

In this article, we develop the framework for directly studying the open quantum systems dynamics from anharmonic oscillator environments that can be explored using numerically exact techniques. This is achieved by utilising a matching criterion developed~\cite{makri1999linear,Makri2024EssentialBathAnharmonicity} for the correlation function of harmonic oscillators and a collection of anharmonic modes in the thermodynamic limit. Using this as a starting point, we derive, for arbitrary uncoupled bath configurations, the correlation functions and, for the first time, study the generic new features that distinguish them from the harmonic case, which has previously not been explored. Furthermore, we develop a framework for studying a few damped anharmonic modes, a regime where the matching criterion fails, and investigate the generic effects of damped anharmonic environments. In Section I, we develop the theoretical framework for calculating the effective harmonic bath spectral density for the mapping of a generic continuum of anharmonic modes to a harmonic one. In Section II, we deploy this framework for an environment of Morse oscillators, where we validate the framework through benchmarking with numerically exact techniques. In addition, this allows for a comparison of the effects of anharmonic and harmonic environments on the system's dynamics. Furthermore, we study the characteristic features that appear within the absorption spectra for anharmonic modes. In Section III, we present a framework for modelling damped anharmonic modes and deriving their associated two-time correlation functions. Leveraging the approach introduced in Section II, we then map the associated two-time correlation function of a single damped anharmonic mode onto an effective continuous bath of harmonic modes. This facilitates a direct comparison of the influences of a single damped anharmonic mode with both the harmonic mapping and a single damped harmonic oscillator on the system's dynamics. We further apply this methodology to explore the influence of anharmonicity on absorption spectra and its capability to enhance energy transport due to the generation of additional environmental resonances. Finally, we extend our analysis to the dynamics of multiple damped anharmonic modes, enabling a systematic comparison with both the continuum and collections of damped harmonic oscillators.

\begin{figure}[ht]
    \centering
    \begin{tikzpicture}
    \node[anchor = south west, inner sep =0]
    (fig) at (0,0)
    {\includegraphics[width=1.0\linewidth]{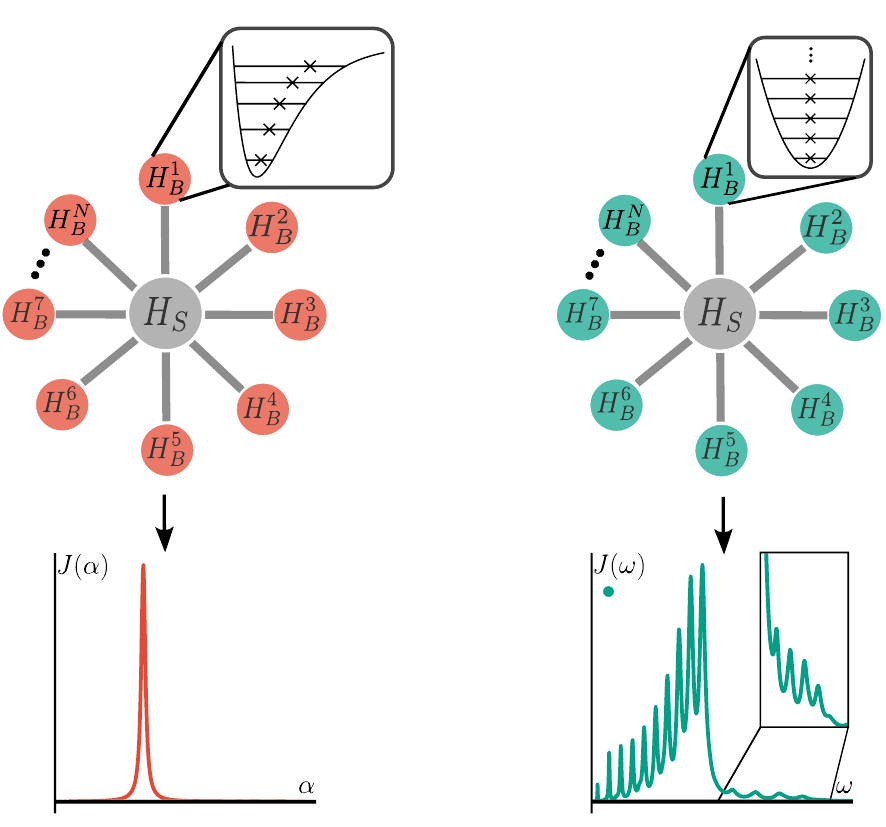}};

    %Text over the figure 
    \node at (0.68,3.5) {$\alpha$-modes};
    \node at (4.2, 4.8) {\fontsize{30}{30}\selectfont=};
    \node at (4.2,5.4) {\small{Harmonic Mapping}};
    \node at (8.2,3.5) {$\omega$-modes};
    \end{tikzpicture}
    \caption{Depiction of the mapping from a continuous anharmonic bath to an equivalent continuous harmonic bath. The grey circles represent the system Hamiltonian, coupled to either red $\alpha$-anharmonic or green $\omega$-harmonic modes, with $H_B^i$ denoting the Hamiltonian of the $i$-th mode. The insert within the red (green) circles illustrates the corresponding potentials, while crosses represent the mean position of each eigenstate $\langle B \rangle$. The bare spectral density produced of the anharmonic modes is represented in red. In the thermodynamic limit of anharmonic modes, $N \rightarrow \infty$, higher-order correlations of the anharmonic bath operator vanish, allowing the bath to be mapped onto an equivalent harmonic environment. This produces an effective spectral density for the harmonic environment with shifted peaks arising from the non-equidistant energy gaps of the anharmonic modes, while its non-zero variance introduces a zero-frequency contribution that acts as static disorder on the system.}
    \label{fig:MorsePotential}
\end{figure}
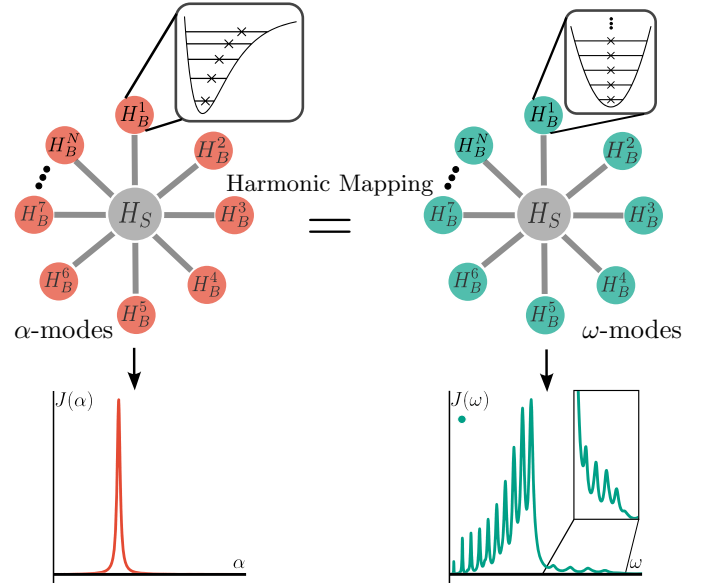

\section{Anharmonic environments}
We consider an open systems model where an environment comprised of a collection of uncoupled anharmonic modes (AMs) couples to an arbitrary system of interest. Such a model is governed by the Hamiltonian 
\begin{equation}
    H = H_S + H_I +H_B, 
\end{equation}
where the system Hamiltonian is $H_S$, the interaction between the anharmonic mode and the system is given by $H_{I}$, and $H_B$ is the free environment Hamiltonian. We begin by assuming that the environment Hamiltonian can be decomposed as a continuum over anharmonic modes, such that it is given by
\begin{equation}
    H_B = \int_0^\infty d\alpha\, H_B(\alpha),
\end{equation}
and by introducing the operators 
\begin{equation}
\overline{\ket{n_{\alpha}}\bra{m_{\alpha}}}\equiv \big(\bigotimes\limits_{\alpha'\neq\alpha } \mathbf{1}_{\alpha'}\big) \otimes \ket{n_\alpha}\bra{m_\alpha},
\end{equation}
that act only in the $\alpha$-mode Hilbert space by transitioning from state $\ket{m_\alpha}\rightarrow\ket{n_\alpha}$, we can write the $\alpha$-mode free Hamiltonian as 
\begin{equation}
    H_B(\alpha)=\sum_{n=0}^{N_\alpha-1}E_{n}(\alpha) \overline{\ket{n_\alpha}\bra{n_\alpha}},
    \label{Eq:Environment_Hamiltonian}
\end{equation}
where $E_{n}(\alpha)$ is the energy of the $n$-th level of the $\alpha$ anharmonic environment mode represented by state $\ket{n_\alpha}$ and $N_{\alpha}$ defines the number of levels of the $\alpha$ anharmonic mode. In addition to denoting an anharmonic mode, $\alpha$ is also used to denote its corresponding frequency in the following sections. For a single system coupling operator $S$, we may write the interaction Hamiltonian in the form 
\begin{equation}
    H_I =\int_0^\infty d\alpha\, H_I(\alpha)=S \otimes \int_0^\infty d\alpha\,g(\alpha) B_\alpha, \label{eqn:IntHam}
\end{equation}
with $g(\alpha)$ the coupling strength of the $\alpha$-mode with bath coupling operator $B_\alpha$.
Following this, as $\big\{\ket{\{n_\alpha\}}\big\}$ provides an orthonormal basis for the environment Hilbert space, we can decompose any arbitrary bath interaction operator for the $\alpha$-mode as
\begin{equation}
    B_\alpha  =\sum_{nm = 0}^{N_{\alpha}-1} B_{nm}(\alpha)\overline{\ket{n_\alpha}\bra{m_\alpha}}.
    \label{Eq:Interaction_Hamiltonian}
\end{equation}
The influence of the environment on the system dynamics can then be characterised, via a cumulant expansion, in terms of $n$-th order correlation functions of the environment coupling operators~\cite{Kubo1962,Kramer2015}. For harmonic environments, all higher-order correlations can be expanded in terms of the two-time correlation function, which fully determines the environmental influence on the system ~\cite{FEYNMAN1963118}. In contrast, for an environment consisting of a few AMs, higher-order correlations cannot be decomposed in such a way. This makes modelling anharmonic environments considerably more challenging~\cite{funo_dynamics_2024, mortiz2022ACE,whitenongaussianbath}. However, as the number of AMs approaches the thermodynamic limit, as is the case in the continuum assumed in this section, the influence of the anharmonic environment is only to modify the two-time bath correlation function, as it becomes Gaussian and thus can be mapped onto a harmonic bath~\cite{makri1999linear}. In Appendix~\ref{App:InfFunc}, a complementary derivation for this has been provided, where we readily see that the $n$-th order cumulants scale as $\mathcal{O}(N^{-n/2+1})$ where $N$ is the number of AMs. This harmonic mapping is achieved by equating the correlation functions of the anharmonic and harmonic environment, enabling the deployment of an effective temperature-dependent spectral density of the form ~\cite{makri1999linear}
\begin{align}
    J_\text{eff}(\omega,\beta) &= \frac{1}{\pi}\tanh (\beta\omega/2) \int^\infty_0 \text{Re}\{\mathcal{C}(t)\}\cos (\omega t) dt\,\label{eqn:effSpec},
\end{align}
with $\beta=1/k_BT$ the inverse temperature of the AM bath, and the correlation function 
\begin{equation}
    \mathcal{C}(t) = \int_0^\infty d\alpha\, |g^2(\alpha)|\langle \mathcal{B}_\alpha(t)\mathcal{B}_\alpha(0)\rangle,
\end{equation}
where the $\mathcal{B}_\alpha=B_\alpha-\langle B_\alpha(0)\rangle$ are the zero-mean AM bath operators. In this article, we use the variable $\alpha$ for anharmonic modes and $\omega$ for harmonic oscillator modes, where $\omega$ denotes the mode's transition energy. The coupling strengths of the effective harmonic environment ($\omega$-modes) are determined through the effective spectral density~\cite{breuer2002theory}. 
It therefore remains to calculate the bath correlation function for an arbitrary AM environment, allowing the corresponding temperature-dependent effective spectral density to be obtained.

\subsection{Single anharmonic mode correlation function}
For simplicity, as we are considering uncoupled environmental modes, we shall first consider the correlation function of a single AM with operator $B_\alpha$. To determine the single-mode AM correlation function defined as 
\begin{equation}
    C(\alpha,t) = \langle B_\alpha(t)B_\alpha(0)\rangle,
\end{equation}
we consider the decomposition of the AM interaction operator in the eigenbasis of its free Hamiltonian $H_{B}(\alpha)$. This yields 
\begin{equation}
    B_\alpha(t) = \sum_{nm=0}^{N_{\alpha} -1}B_{nm}(\alpha)e^{iE_{ nm}(\alpha)t}\overline{\ket{n_\alpha}\bra{m_\alpha}},
\end{equation}
with $E_{ nm}(\alpha) = E_{ n}(\alpha)-E_{ m}(\alpha)$.
For an initial Gibbs state of the bath mode 
\begin{equation}
    \rho_\alpha^\beta =\sum_{n=0}^{N_\alpha -1}  p_n(\alpha)\overline{\ket{n_\alpha}\bra{n_\alpha}}, 
\end{equation}
where $p_n(\alpha) = e^{-\beta E_{ n}(\alpha)}/Z(\alpha)$ is the probability of being in state $\ket{n_\alpha}$, and the partition function is given by $Z(\alpha)= \sum_{n=0}^{N_{\alpha}-1} e^{-\beta E_{ n}(\alpha)}$. The bath correlation function can then be decomposed as 
\begin{align}
    C(\alpha,t) &= \Tr{B_\alpha(t)B_\alpha(0)\rho^\beta_\alpha}\nonumber\\ &=\Tr\bigg\{\sum_{nmklr = 0}^{N_{\alpha}-1} B_{nm}(\alpha)e^{iE_{ nm}(\alpha)t}\overline{\ket{n_\alpha}\bra{m_\alpha} } \nonumber\\ & \hspace{40pt}\times B_{kl}(\alpha)\overline{\ket{k_\alpha}\bra{l_\alpha}}p_{r}(\alpha)\overline{\ket{r_\alpha}\bra{r_\alpha}}\bigg\}\nonumber\\&=\sum_{n,m = 0}^{N_{\alpha} -1} |B_{nm}(\alpha)|^{ 2}e^{iE_{nm}(\alpha)t}p_n(\alpha).
    \label{CorrFuncC(alpha,t)}
\end{align}
It is useful to decompose any such AM coupling operator into diagonal and off-diagonal components as
\begin{align}
    B_\alpha &= D_{\alpha } + O_{\alpha },\\
    D_{\alpha } &= \sum_{n=0}^{N_{\alpha} -1} B_{nn}(\alpha)\overline{\ket{n_\alpha}\bra{n_\alpha}},\\
    O_{\alpha } &= \sum_{n\neq m = 0}^{N_{\alpha} -1} B_{nm}(\alpha)\overline{\ket{n_\alpha}\bra{m_\alpha}}.
\end{align}
Note that in the interaction picture, the diagonal terms in $D_{\alpha }$ do not acquire a time-evolving phase $D_{\alpha }(t)=D_{\alpha}(0)$. It is convenient for the following analysis to assume that the coupling operator has zero mean, which can always be achieved by performing a simple displacement of the bath operator 
\begin{equation}
    \mathcal{B}_{\alpha} = B_\alpha -\langle B_\alpha\rangle 
    \label{Eq:appropriate_coupling}, 
\end{equation}
where $\langle B_\alpha \rangle = \Tr\{B_\alpha\rho^\beta_\alpha\}$ such that $ \langle\Tilde{B}_\alpha\rangle=0$. In order to leave the total Hamiltonian unchanged, this shift is added to the system Hamiltonian
\begin{equation}
    {H}_S  \rightarrow H_S+  S\langle B_\alpha\rangle.
\end{equation}
Using the eigenbasis decomposition, with a Gibbs state for the environment, the linear moment of the bath operator is solely determined by the diagonal components of $B$ as 
\begin{equation}
 \langle B_\alpha \rangle = \sum_{n = 0}^{N_{\alpha} -1 } B_{nn}(\alpha)p_n(\alpha)  = \langle D_{\alpha}\rangle.
\end{equation}
We can then calculate the correlation function for the displaced operator as 
\begin{equation}
    \mathcal{C}(\alpha,t) = \langle \mathcal{B}_\alpha(t)\mathcal{B}_\alpha(0)\rangle =  C(\alpha,t) - \langle B_\alpha\rangle^2,
\end{equation}
and using the derivation of the original AM correlation function, Eq.~\eqref{CorrFuncC(alpha,t)}, we find a form,
\begin{align}
    \mathcal{C}(\alpha,t) &=\sum_{n\neq m = 0}^{N_{\alpha} - 1} |B_{nm}(\alpha)|^2e^{iE_{ nm}(\alpha)t}p_n(\alpha) \nonumber\\
    &+ \sum_{n= 0}^{N_{\alpha} -1} B_{nn}(\alpha)^2p_n(\alpha)- \bigg(\sum_{n = 0}^{N_{\alpha} -1}B_{nn}(\alpha)p_n(\alpha)\bigg)^2 \nonumber\\&= C_{O}(\alpha,t) + \Delta D_\alpha^2\label{eqn:TildeCorr}.
\end{align}
This is equivalent to the correlation function of the off-diagonal operator
\begin{equation}
    C_{O}(\alpha,t) = \langle O_\alpha(t)O_\alpha(0)\rangle,
\end{equation}
plus the variance of the diagonal operator
\begin{equation}
    \Delta D_\alpha^2 = \langle D_\alpha^2 \rangle -\langle D_\alpha\rangle^2.
\end{equation}
This leads to a non-zero time-averaged value for the single AM correlation function when the variance of the diagonal operator $D_\alpha$ is non-zero.
\subsection{Effective spectral density}
The above results were for a single AM; due to the AMs being uncoupled, the total environment correlation function can be constructed by integrating over these modes
\begin{align}
    &\mathcal{C}(t) = \int_0^\infty d\alpha\, g^2(\alpha) \mathcal{C}(\alpha,t)\label{eqn:Corr}\\
    &=\int_0^\infty d\alpha\, J(\alpha)\bigg(\sum_{n\neq m}^{N_{\alpha} -1} |B_{nm}(\alpha)|^2e^{iE_{ nm}(\alpha)t}p_n(\alpha)+\Delta D_\alpha^2\bigg),\nonumber
\end{align}
where $J(\alpha) = g^{2}(\alpha)$. We now use Eq.~\eqref{eqn:Corr} and Eq.~\eqref{eqn:effSpec}, where, for simplicity, we assume that all the AMs have the same energy level up to linear scaling by $\alpha$ (i.e. $E_{nm}(\alpha) = \alpha E_{nm}$). Appendix~\ref{App:ContSpecDens} contains the solution without this assumption, which takes a similar form. We calculate for the first time the AM environment's effective harmonic spectral density as 
\begin{align}
    J_\text{eff}&(\omega,\beta) =\frac{1}{\pi}\tanh (\beta\omega/2) \int^\infty_0 \text{Re}\{\mathcal{C}(t)\}\cos (\omega t) dt\,\nonumber\\ =&\frac{1}{2}(1-e^{-\beta\omega})\times\nonumber\\ &\sum_{n>m = 0}^{N_{\alpha} -1} |B_{nm}\bigg(\frac{\omega}{E_{nm}}\bigg)|^2J\bigg(\frac{\omega}{E_{nm}}\bigg)
    \frac{1}{E_{nm}}p_n\bigg(\frac{\omega}{E_{nm}}\bigg) \nonumber\\&\hspace{20pt}+A_\text{v}\delta(\omega),
    \label{eqn:effective_spectral_density}
\end{align}
where
\begin{equation}
A_\text{v}=\int_0^\infty d\alpha\, J(\alpha)\Delta D^2_\alpha
\label{eqn:non_zero_variance}
\end{equation}
is the contribution from the non-zero steady state of the correlation functions due to the variance of the diagonal AM coupling operators $D_\alpha$ at equilibrium and $E_{nm} = \alpha/E_{nm}(\alpha)$. As above, we use $\omega$ to denote the harmonic frequency modes with energy $\omega$ to distinguish them from $\alpha$ for anharmonic modes. 

Firstly, we note that for a harmonic oscillator system with linear bath coupling ($B_\alpha=x_\alpha$), the above equation yields the underlying spectral density $J_\text{eff}(\omega,\beta)=J(\omega)$ as the position operator has no diagonal component $x_{nn}(\alpha)=0$, thus $D_\alpha=0$, due to the symmetry of the harmonic potential. Furthermore, we note that in the AM case --- where adjacent energy levels are not equally spaced --- we have an effective representation comprised of a summation of shifted spectral densities, shifted by the relative transition $\omega/E_{nm}$ and weighted by the thermal Gibbs weights. The last term, $A_\text{v}\delta(\omega)$, associated with the variance of the diagonal operator, is a unique feature of the AM bath. Superficially, this term appears problematic for accurate modelling within the framework of open quantum systems due to the $\delta$ feature. However, to match the bath correlation function in the effective harmonic framework, we may simply extend our `system' Hilbert space by the inclusion of a single harmonic oscillator with zero transition energy coupled to the system with coupling strength $g = \sqrt{A_\text{v}}$, initially in its ground state. This enables the removal of this divergence from the spectral density, and the residual `well-behaved' spectral density can be considered, utilising conventional approaches for modelling harmonic oscillator environments ~\cite{chin_exact_2010, mortiz2022ACE}. It should be noted that this is equivalent to modelling static disorder with a strength governed by $A_\text{v}$, implying that the variance in the AMs diagonal operator acts as a classical static noise channel on the system. This is further explored in its effects on the absorption spectrum.

Taking the mapping a step further, the thermalised spectral density framework shows that non-zero temperature Gaussian harmonic oscillator environments have an equivalent representation as a zero-temperature harmonic oscillator environment with frequencies extended to the negative reals and temperature-dependent couplings~\cite{tamascelli_efficient_2019}. This enables the AM environment framework to be deployed with open quantum systems methods that require zero-temperature environments or are purely wavefunction-based. The thermalised spectral density has the form
\begin{align}
    J_\text{eff}^\text{th}(\omega,\beta) &= J_\text{eff}(\omega,\beta)(n(\omega,\beta)+1)\Theta(\omega) \nonumber\\&+ J_\text{eff}(-\omega,\beta)n(-\omega,\beta)\Theta(-\omega)\nonumber\\
    &=\frac{1}{\pi}\gamma(\omega),
    \label{Eq:thermalisedspectraldensity}
\end{align}
where the rate function is given by
\begin{align}
    \gamma(\nu) &=\Re{\int_0^\infty dt\, e^{-i\nu t}\mathcal{C}(t)}\nonumber\\
    &=\begin{cases}
        \pi\sum_{n>m = 0}^{N_{\alpha} -1}\frac{1}{E_{nm}}X_{nm}(\frac{\nu}{E_{nm}}) \text{ for }\nu >0 \\
        \pi\sum_{n<m = 0}^{N_{\alpha} -1}\frac{1}{|E_{nm}|}X_{mn}(\frac{\nu}{E_{nm}}) \text{ for }\nu <0
    \end{cases},
\end{align}
with 
\begin{equation}
    X_{nm}(\alpha)= J(\alpha) |B_{nm}(\alpha)|^2p_n(\alpha).
\end{equation}
We use $\nu$ to denote a frequency, and $\gamma(\nu)$ is the frequency response of the correlation function. By construction, the harmonic oscillator thermalised effective spectral density is equal to the frequency response of the AM bath correlation function; see Appendix~\ref{App:ContSpecDens} for full details. 
This maps the non-zero temperature anharmonic modes to a zero temperature harmonic oscillator environment with modified spectral density given by $J^\text{th}_\text{eff}(\omega,\beta)$. $n(\omega,\beta)= 1/(e^{\beta\omega}-1)$ is the average occupation of the harmonic oscillator given by the Bose-Einstein distribution. In the following sections, we consider the Morse oscillator as a specific emblematic example of anharmonicity and outline the effectiveness of the developed approach, as well as some of the interesting, unique features within Morse oscillators.

\subsection{Absorption spectra}
\label{Absorption_Spectra}
The development of spectroscopic techniques has enabled the direct probing of the energy structure and dynamical response of quantum systems~\cite{Jonas2003,Engel2007, Roos, lorenzoni_persistent_2025}. These measurements provide detailed access to system-environment interactions, motivating the development of accurate theoretical and numerical methods for interpreting the environment interactions. In this context, absorption spectra are particularly important as they encode both the intrinsic properties of the system and the influence of the environment, allowing a probe into anharmonic effects ~\cite{WangThoss2004, Smith2019QuantumDissipative}. Within the independent spin boson model, one can analytically calculate the absorption spectra of the system via the Fourier transform of the optical coherence function 
\begin{equation}
    \theta(t) = \exp[\Gamma(t)-\Gamma(0)]/2,
\end{equation}
where $\Gamma(t)$ is the lineshape function describing the dephasing of the optical coherence induced by the system–environment interaction ~\cite{deVega2017}
\begin{equation}
    \Gamma(t) = \int_0^\infty d\omega\, \frac{J_\text{eff}(\omega,\beta)}{\omega^2} (\coth\big(\frac{\beta\omega}{2}\big)\cos\omega t +i\sin\omega t). 
\end{equation}
One can substitute the effective spectral density, Eq.~\eqref{eqn:effective_spectral_density}, to find that the general form of the lineshape function for anharmonic systems is 
\begin{align}
  \Gamma&(t) =  \frac{-A_\text{v} t^2}{2} \nonumber\\&\hspace{10pt}+ \int_0^\infty d\alpha\, J(\alpha)\sum_{n\neq m}^{N_{\alpha}-1} p_n(\alpha) |B_{nm}(\alpha)|^2 \frac{e^{-iE_{ nm}(\alpha)t}}{E_{nm}(\alpha)^2}\nonumber\\
  &= \Gamma_s(t) + \Gamma_d(t),
  \label{lineshpe_function}
\end{align}
where $\Gamma_{s,d}(t)$ corresponds to the lineshape function for the static and dynamic parts of the environment. 
From this, we can immediately see that the coherence function has a Gaussian suppression due to the variance of the diagonal bath operator $D_\alpha$ with 
\begin{align}
        \theta(t) &= \frac{e^{-\frac{A_\text{v}t^2}{2}}}{2} \times \nonumber\\&\hspace{0pt}\exp\bigg\{\int_0^\infty d\alpha\, J(\alpha)\sum_{n\neq m }^{N_{\alpha} -1} p_n(\alpha) \nonumber\\&\hspace{50pt}\times|B_{nm}(\alpha)|^2 \frac{e^{-iE_{nm}(\alpha)t}-1}{E_{\alpha, nm}^2}\bigg\}\nonumber\\&=\theta_\text{s}(t) \theta_{\text{d}}(t),
        \label{Eq:DecayofCoherences}
\end{align}
where $\theta_{\text{s,d}}$ are the contributions to the coherence function from the static and dynamical parts of the environment corresponding to $D_\alpha$ and $O_\alpha$, respectively. 
The absorption spectrum is the Fourier transform of the coherence function
\begin{align}
    A(\omega) &= \mathcal{F}\{\theta(t)\}(\omega)\nonumber \\&= \mathcal{F}\{\theta_s(t)\}*\mathcal{F}\{\theta_d(t)\}\nonumber\\&=A_s(\omega)*A_d(\omega),
\end{align}
where $*$ denotes a convolution and $A_{s,d}$ are the absorption spectra from the static and dynamical components of the environment.  This indicates that the variance parameter acts much like inhomogeneous line broadening due to static disorder, as it convolves our dynamical part (homogeneous term) with a Gaussian profile.  One can quantify the relative inhomogeneous/homogeneous line broadening by the ratio of the fluctuations they generate, recalling that the variance term arises from the diagonal part of the coupling operator and the dynamical term from the off-diagonal components. This ratio takes the form
\begin{equation}
    R = \frac{\int_0^\infty d\alpha\, J(\alpha)\Delta D^2_\alpha}{\int_0^\infty d\alpha\, J(\alpha)\Delta O^2_\alpha}.
\end{equation}
Expanding the spectral function to first order in the dynamic lineshape function, we find that 
\begin{equation}
    A_d(\omega) \propto \pi\kappa\bigg(\delta(\omega)+ \frac{ J_\text{eff}^\text{th}(\omega, \beta) }{\omega^2}\bigg),
\end{equation}
where $\kappa = \exp\{-\Gamma(0)\}$ and the $\delta$-function corresponds to the zero phonon-line and $J_\text{eff}^\text{th}(\omega)/\omega^2$ is the single AM contribution. We can calculate the absorption spectrum due to the static variance noise as
\begin{equation}
    A_s(\omega) \propto \exp\{-\frac{\omega^2}{2A_\text{v}}\},
\end{equation}
yielding the total spectrum 
\begin{equation}
    A(\omega) \propto \pi\kappa\bigg( \exp\{-\frac{\omega^2}{2A_\text{v}}\} + \exp\{-\frac{\omega^2}{2A_\text{v}}\}*\frac{J_\text{eff}^\text{th}(\omega,\beta)}{\omega^2}\bigg).
\end{equation}
This suggests that the AMs introduce a characteristic feature into the absorption spectrum: A central Gaussian profile that captures the variance in the diagonal component of the coupling operator. It should be made explicit that this effective static disorder arises solely from unitary dynamics and not from the addition of any phenomenological noise. Moreover, by measuring the width of this central spectrum, we may determine the variance parameter $A_\text{v}$. Furthermore, for lower temperatures, such that $A_\text{v}\approx 0$, the broadening associated with the central spectrum becomes negligible. Nevertheless, for appreciable temperatures that still ensure thermal occupation of excited vibrational states, we anticipate that the absorption spectrum should contain clear features of the anharmonicity through the effective spectral density $J_\text{eff}^\text{th}(\omega,\beta)$. We demonstrate the absorption spectra for a Morse oscillator bath in Section III B. 

\section{Morse potential}
\label{SectionII}

\begin{figure*}[ht]
    \centering
    \includegraphics[width = \linewidth]{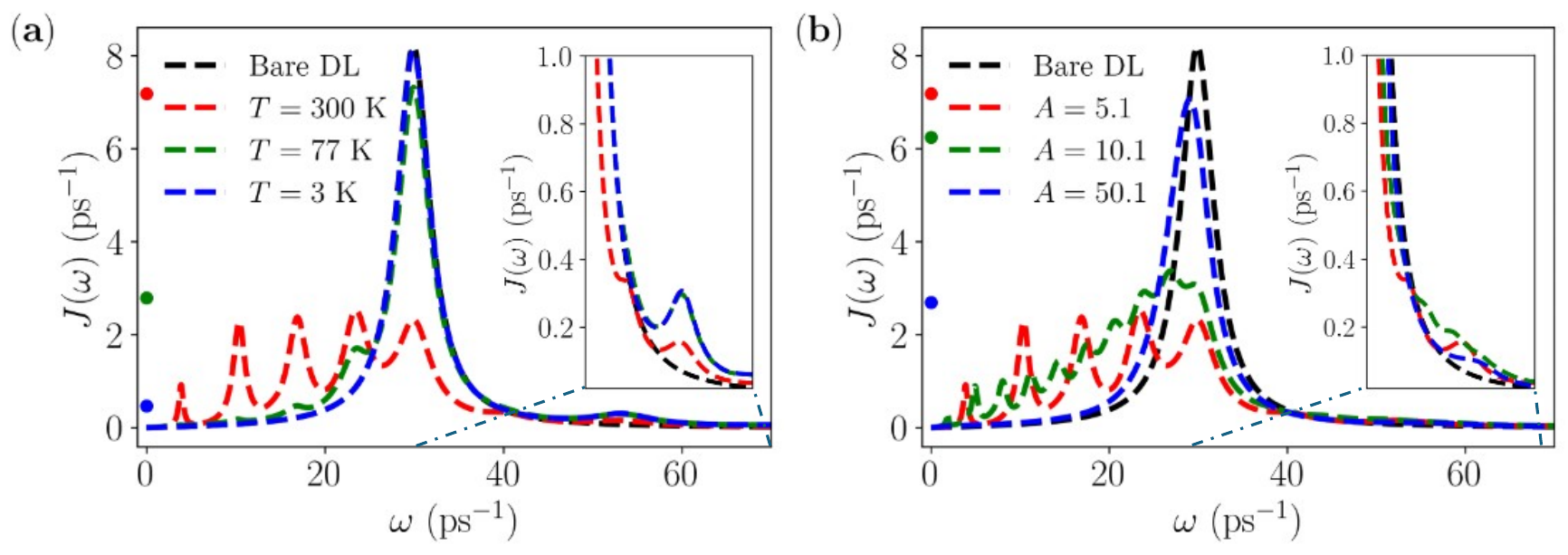}
        \caption{\label{fig:morse_eff_SD} The effective spectral density of a Morse oscillator environment with (a) varying temperatures for the ensemble with anharmonicity parameter $A=5.1$ and (b) varying anharmonicity parameter $A$ at 300~K, including the coupling of the zero-frequency term indicated at zero frequency. The underlying spectral density is the underdamped Ohmic Lorentzian defined as Bare DL (Eq.~\eqref{eqn:underdampedSD}) with parameters $\Omega = 30$~ps$^{-1}$, $\gamma=4.5$~ps$^{-1}$ and $\lambda = 0.01$~ps$^{-1}$. }
    \label{Fig:Effective_Spectral_Density}
\end{figure*}
In this section, we consider a physically motivated anharmonic environment, a collection of Morse oscillators. The Morse potential is widely used to model the interatomic potential energy of diatomic molecules; this can provide a more realistic description of molecular vibration than the quantum harmonic oscillator when anharmonic effects become significant. Unlike the harmonic potential, the Morse potential captures the finite dissociation energy of chemical bonds and exhibits non-uniform vibrational level spacings with a finite number of bound states. These features become important when higher vibrational levels are thermally populated. Conversely, when the vibrational mode remains predominantly in its ground state, such as when the oscillator frequency is large compared with the thermal energy of the bath, the harmonic approximation is expected to provide an accurate description. The exact Morse spectrum has been calculated ~\cite{delima2005morse}, allowing for analytical calculation of the effective spectral density for an environment comprised of Morse oscillators. The Morse potential has the compact form
\begin{equation}
    V(r) = D_e(1-e^{-a(r-r_e)})^2,
\end{equation}
where $D_e$ is the classical dissociation energy, $r_e$, the equilibrium position and $a$ determines the potential width. 
Such a potential leads to a collection of discrete bound states and a continuum of unbound states above the dissociation energy, where bond breaking occurs. The discrete bound states of a Morse oscillator are shown in Fig.~\ref{fig:MorsePotential}. Here, we consider only the bound states of the Morse oscillator, as exploring the thermodynamics of the continuum of unbound states is beyond the scope of this work~\cite{ENMiranda_2001}. Defining the dimensionless Morse parameter $\lambda = \frac{\sqrt{2MD_e}}{a\hbar}$, where $M$ is the mass of the particle, the bound state energy spectrum has a quadratic form 
\begin{equation}
    H_B(\alpha) \ket{n_\alpha} = -\alpha D_e(A-n)^2\ket{n_\alpha},
\end{equation}
where $A =\lambda-\frac{1}{2}$ allowing us to define $E_{\alpha,n} = -\alpha D_{e}(A - n)^{2}$. Such a system exhibits $\lfloor A+1\rfloor$ bound states with $n\in\{0,1,...,\lfloor A\rfloor\}$. Furthermore, the position operator $x = a(r-r_e)$ for the discrete states can be written as 
\begin{equation}
    x_\alpha=  \sum_{n,m =0}^{\lfloor A \rfloor }x_{nm}\ket{n_\alpha}\bra{m_\alpha}, 
\end{equation}
assuming that the coefficients are independent of frequency, for $m>n$
\begin{align}
    x_{nm} = &  \frac{2(-1)^{m-n+1}}{(m-n)(2A-n-m)}\\ &\times \sqrt{\frac{(A-n)(A-m)\Gamma(2A-m+1)m!}{\Gamma(2A-n+1)n!}},\nonumber
\end{align}
where $\Gamma$ is the gamma function.  For the diagonal terms
\begin{align}
    x_{nn} = &\ln(2A+1) +\psi(2A-n+1)\nonumber\\&-\psi(2A-2n+1)-\psi(2A-2n),
\end{align}
 where $\psi$ is the digamma function.
These diagonal terms lead to $x_\alpha$ having a non-zero linear moment $\Bar{x}_\alpha=\langle x_\alpha\rangle$ as seen in Fig.~\ref{fig:MorsePotential}, where the crosses indicate the average position of each eigenstate. One can calculate the partition function for a single Morse oscillator as a simple partial sum
\begin{equation}
    Z_\alpha = \sum_{n=0}^{\lfloor A\rfloor} e^{\beta \alpha D_e(A-n)^2},
\end{equation}
Thus, the associated probability for being in the $\ket{n}_\alpha$ state is 
\begin{equation}
    p_n(\alpha) = \frac{e^{\beta \alpha D_e(A-n)^2}}{Z_\alpha}.
\end{equation}
This then leads to the linear moment of the coupling operator
\begin{equation}
    \bar{x}_\alpha= \sum_{n=0}^{\lfloor A \rfloor} x_{nn}p_n(\alpha).
\end{equation}

Utilising the framework developed above, and assuming that our system couples to the Morse oscillators through their position operators
\begin{equation}
    H_I  = S\otimes \int d\alpha\, g(\alpha) x_\alpha,
\end{equation}
we can calculate the effective spectral density as 
\begin{align}
    &J_\text{eff}(\omega,\beta) =(1-e^{-\beta\omega})     \bigg[A_{v}\delta(\omega) + \\& \sum_{n<m}^{\lfloor A \rfloor} \frac{(2A-1)}{(m-n)(2A -n-m)}J(\omega_{nm}) x_{nm}^{2} p_n(\omega_{nm}) \bigg],\nonumber
\end{align}
where we have defined $\omega_{nm} = \frac{\omega(2A-1)}{(m-n)(2A-n-m)}$. The variance associated with the diagonal part of the coupling operators can thus be written as 
\begin{equation}
    V_x(\alpha) = \sum_{n=0}^{\lfloor A \rfloor} x_{nn}^2p_n(\alpha)- \left(\sum_{n=0}^{\lfloor A \rfloor} x_{nn}p_n(\alpha)\right)^2.
\end{equation}
We can then calculate the effective coupling strength of the zero-frequency oscillator by integrating the spectral density
\begin{equation}
    g_0^2= \int_0^\infty d\alpha\, J(\alpha)V_x(\alpha).
\end{equation}

In this framework, we map the effects of the anharmonic environment to an effective harmonic one; the effects of anharmonicity appear as additional structure in the spectral density. To expose this, we consider a bare underlying spectral density that has the underdamped Ohmic Drude-Lorentz form,
\begin{equation}
    J(\omega) = \frac{4\lambda\omega(\Omega^2+\gamma^2) \gamma \omega}{(\omega^2-\Omega^2)^2+\omega^2\gamma^2}.
    \label{eqn:underdampedSD}
\end{equation}

The results for the effective spectral densities are shown in Fig.~\ref{Fig:Effective_Spectral_Density}. Fig.~\ref{Fig:Effective_Spectral_Density}(a) demonstrates the temperature dependence of the effective spectral densities. When the thermal energy $k_{B}T$ becomes comparable to the energy of the anharmonic mode $E_{\alpha, nm}$, distinct anharmonic effects become apparent as higher-energy states of the mode become thermally occupied. This leads to the emergence of multiple peaks in the effective spectral density, resulting in a broadening and a reduction in the effective coupling strength at the bare resonance frequency $\Omega$. The shifted peaks arise from the unequal level spacing of the anharmonic mode, which produces a structured effective spectral density with resonances at the AM transition energies. The Morse oscillator position operator enables transitions beyond adjacent energy levels; this results in red- and blue-shifted features within the effective spectral density, such that larger (smaller) energy gaps than the ground to first excited state lead to blue (red) shifted peaks. Another key feature of the Morse oscillator ensemble is the zero-frequency mode and its coupling strength to the system, which, in the effective spectral density, is represented as a single point at $\omega = 0$. We find that the zero-frequency contribution increases with both temperature and anharmonicity of the Morse ensemble, as higher excited states become increasingly populated, and correspondingly the variance of oscillators' positions grows. In Fig.~\ref{Fig:Effective_Spectral_Density}(b), we see the effects of tuning the anharmonicity parameter $A$ associated with the potential depth of the Morse oscillators at $T = 300$~K, modifying the number of bound states. Similar to the temperature effects, we note that the potential depth, which determines the energy level structure, moderates the anharmonic effects seen within the effective spectral density produced through duplicated peaks, enabling rich resonant conditions to become exploitable, as seen in Fig.~\ref{Fig:Effective_Spectral_Density}(b) for low potential depth. Furthermore, we see in Fig.~\ref{Fig:Effective_Spectral_Density}(b) that as we decrease anharmonicity (larger $A$ parameter), features become less distinct as the energy level spacing in the Morse oscillators approaches equal spacing, and the potential becomes harmonic, reproducing the bare spectral density. Another interesting feature of the Morse oscillator's effective spectral density is its enhanced Ohmicity near the origin. The Ohmicity $s$ is defined by the leading-order term in the Maclaurin expansion,
$J(\omega)=\alpha\omega^s+\mathcal{O}(\omega^{s+1})$. Examining the low-frequency limit, we find that within this model $p_n(\omega)\rightarrow 1/\lfloor A+1\rfloor$. Consequently, the effective spectral density acquires an additional linear frequency dependence, increasing the Ohmicity parameter by one from the $(1-e^{-\beta\omega})$ term.
\begin{figure*}[ht]
    \centering
    \includegraphics[width = \linewidth]{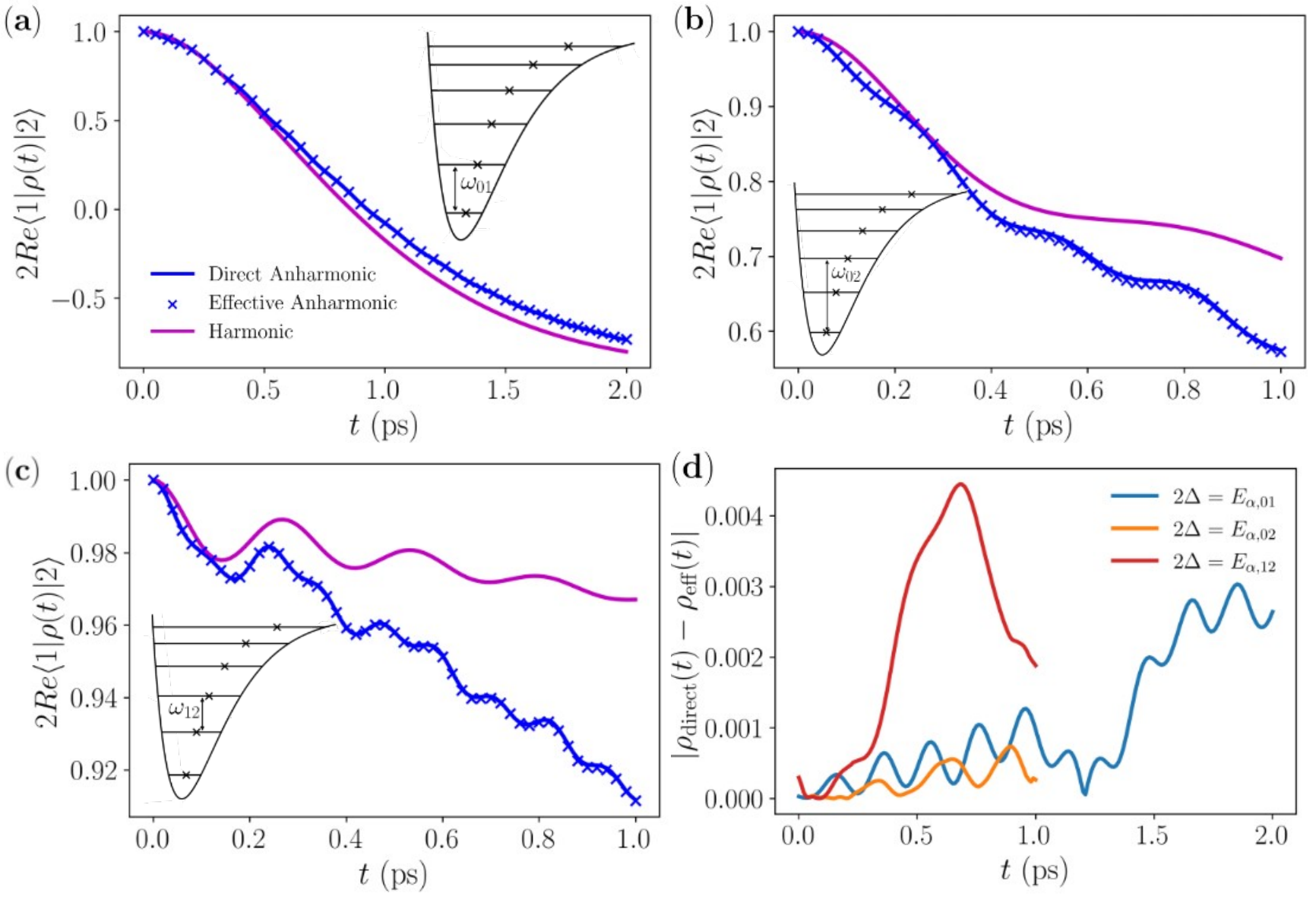}
    \caption{\label{fig:sx_dynamics_validation} Coherence dynamics of a dimer system coupled to a Morse potential-anharmonic environment with a potential depth of $A = 5.1$ and an underlying underdamped spectral density given by Eq.~\eqref{eqn:underdampedSD}. The direct approach models the environment as a collection of anharmonic oscillations referred to as `direct anharmonic', whereas `effective anharmonic' and `harmonic' model the environment as a collection of harmonic oscillators with the effective and bare spectral densities, respectively. The spectral density parameters are $\Omega = 30$~ps$^{-1}$, $\gamma = 0.15 \Omega$, $\lambda = 0.001$~ps$^{-1}$, with system paramaters $\epsilon_{0} = \epsilon_{1} = 0$ and coupling between monomers are $\Delta=E_{\alpha, 01}/2$, $\Delta=E_{\alpha,12}/2$, and $\Delta=E_{\alpha,02}/2$ for panels (a), (b), and (c), respectively. The insert indicates the transitions the system is resonant with in each case.  (d) Plots the absolute difference across time between the direct anharmonic and the effective anharmonic for each distinct case. }  
\end{figure*}

\begin{figure}[ht]
    \centering
    \includegraphics[width = \linewidth]{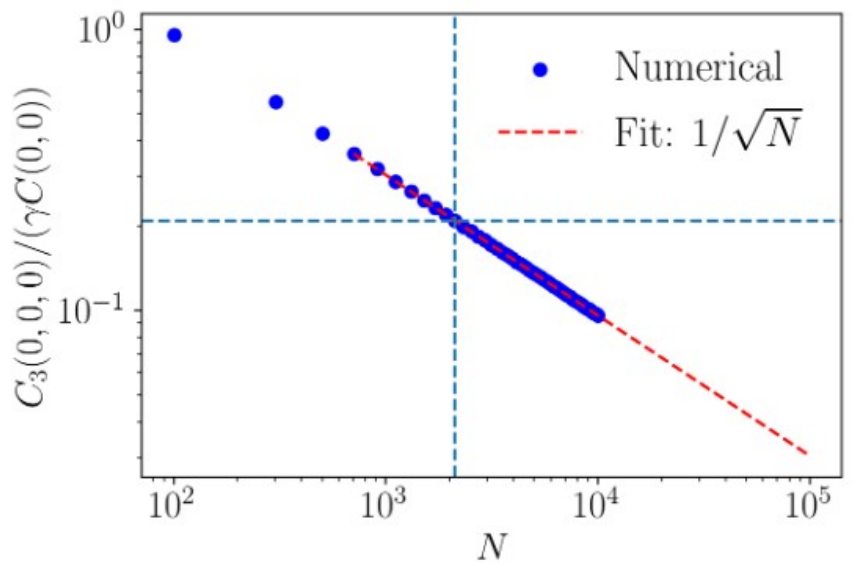}
    \caption{\label{fig:Third_Order_Correlation_Function} Heuristic estimate for the number of discretised modes $(N)$ required for the ratio of the third-order correlation function $C_3$ and second-order correlation function $C$ for a Morse oscillator bath with spectral density defined in Fig.~\ref{Fig:Effective_Spectral_Density} to tend to zero. The intersection of the dashed lines indicates the number of modes used in the direct ACE simulations of the dimer coherence dynamics. This demonstrates that the discrepancy between the direct and effective mapping approaches arises from the finite discretization of the environment.}  
\end{figure}

\subsection{Dynamics of a dimer coupled to a Morse oscillator ensemble}

\begin{figure*}[ht]
    \centering
    \includegraphics[width = \linewidth]{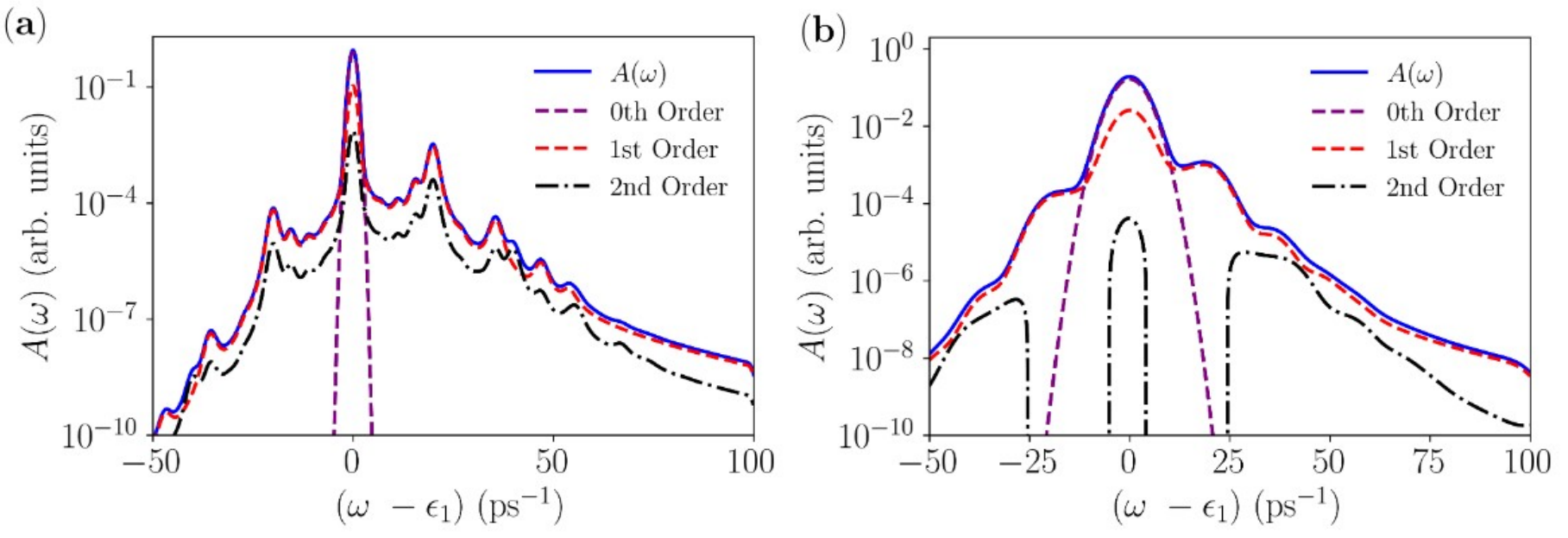}
    \caption{\label{fig:absorption_spectra_continuous} Absorption spectra for a molecular dimer within the single-excitation manifold for anharmonicity parameter $A = 5.1$ at (a) $T = 20$~K and (b) $T = 77$~K with spectral density parameters $\gamma = 1.0$~ps$^{-1}$, $\lambda = 0.001$~ps$^{-1}$ and $\Omega = 20$~ps$^{-1}$. Within each plot, the phonon processes within the absorption spectra are plotted up to $2$nd order.}  
\end{figure*}
To validate the approach, we utilise numerically exact techniques to calculate the dynamics of a monomer system coupled to an environment of Morse oscillators, which we call the direct anharmonic approach, and the effective anharmonic approach, in which the continuum of anharmonic modes is mapped onto a harmonic environment using the results from the previous sections. To achieve numerically exact dynamics, we use the automatic compression of environments (ACE) method ~\cite{mortiz2022ACE}. ACE can be used to model both anharmonic and harmonic environments as it does not rely on the environment being harmonic or Gaussian. Instead, it only requires that the environment can be decomposed into a set of individual degrees of freedom. The resulting system dynamics can then be simulated efficiently using tensor-network techniques by combining and compressing the individual environment degrees of freedom into the relevant influence on the system's dynamics. This approach enables the simulation of systems coupled to structured environments ~\cite{wiercinski_role_2024}.

The model system we consider is a molecular dimer within the single-excitation manifold. The two-level system gives the associated Hamiltonian,
\begin{equation}
    H_{S} = \epsilon_{1}\ket{1}\bra{1} + \epsilon_{2}\ket{2}\bra{2} + \Delta(\ket{1}\bra{2} + \ket{2}\bra{1}).
    \label{Eq.Molecular_Dimer}
\end{equation}
$\epsilon_i$ denotes the energy of the $i$th monomer, and $\Delta$ represents the coupling between monomers that may arise from dipole–dipole interactions. The system couples to the environment via the operator $S =(\ket{1}\bra{1}-\ket{2}\bra{2})$, with the bath at a temperature of $T = 77$~K and described by a damped Ohmic Lorentz–Drude spectral density (Eq.~\eqref{eqn:underdampedSD}), with parameters specified in Fig.~\ref{Fig:Effective_Spectral_Density}. The Morse anharmonicity parameter is set to $A = 5.1$ such that anharmonic effects are visible in the effective spectral density as shown in Fig.~\ref{Fig:Effective_Spectral_Density}(a). We consider three distinct coupling regimes and analyse the resulting coherent dynamics via the coherence $2\Re{ \bra{1}\rho\ket{2}}$. We choose $\Delta_{1} = E_{\alpha,01}/2$, $\Delta_{2} = E_{\alpha,02}/2$, and $\Delta_{3} = E_{\alpha,12}/2$, corresponding to different resonances of the Morse oscillator environment with ground to excited state energy given by $\alpha = 30$~ps$^{-1}$, which corresponds in the spectral density to the resonant, first blue- and red-shifted peaks respectively, as indicated in the insets of Fig.~\ref{fig:sx_dynamics_validation}(a–c). In all cases, the monomer energies are degenerate ($\epsilon_{1} = \epsilon_{2}$). For each resonant case, we validate the mapping by comparing the coherence dynamics of the system when coupled to the effective anharmonic environment with the direct anharmonic environment. Furthermore, the coherence dynamics obtained for a harmonic environment are presented, allowing for a comparison of the additional non-Markovian effects of anharmonic environments. To isolate the intrinsic effects of anharmonicity, we choose the coupling operator to be the displaced position operator $B_\alpha=x_\alpha-\langle x_\alpha \rangle$.

First, we observe that the direct and effective anharmonic approaches yield the same dynamics across all three cases, as shown in Fig.~\ref{fig:sx_dynamics_validation}(d), with absolute errors on the order of $10^{-3}$. The error primarily arises from the direct anharmonic approach. In ACE, the environment must be discretised into individual degrees of freedom to simulate the influence of the environment on the system. Consequently, the direct approach requires an increasingly large number of environmental modes to accurately approach the thermodynamic limit. Fig.~\ref{fig:Third_Order_Correlation_Function} provides a heuristic for this convergence by showing the relative magnitude of the third-order contributions to the second order for this anharmonic environment, which decreases towards zero as the number of environmental modes is increased; these third-order contributions are expected to decay as $\mathcal{O}(N^{-1/2})$, as is evidenced by the fitted curve. However, since ACE is a numerically exact tensor-network technique, increasing the discretization to $\sim 10^{4}$ modes becomes computationally prohibitive. The discrepancy between the direct and effective approaches can therefore be attributed to the small but finite contribution of higher-order correlations arising from the finite discretisation of the environment. When the system transition energy ($2\Delta$) is resonant with the first transition of the Morse oscillator- corresponding to the central peak in the spectral density- the harmonic environment induces stronger decoherence than the anharmonic environment. This originates from the reduced system-environment coupling in the effective spectral density $(J_{\textbf{eff}}(\omega, \beta))$ relative to the bare spectral density $(J(\omega))$ as seen within Fig.~\ref{Fig:Effective_Spectral_Density}(a). Conversely, when the system is resonant with the red- or blue-shifted peaks, as demonstrated within the inserts of Fig.~\ref{fig:sx_dynamics_validation}(b-c), the effective spectral density exhibits stronger coupling compared to the bare spectral density, as shown in Fig.~\ref{Fig:Effective_Spectral_Density}(a). This results in the anharmonic environment inducing stronger decoherence than the harmonic approximation, leading to notable differences in dynamics. These results highlight that, compared to the harmonic assumption, the anharmonic environment induces additional non-Markovian effects that can either enhance or diminish decoherence of the system. 

\subsection{Absorption spectrum for a monomer coupled to a Morse oscillator ensemble}
\label{Absorption_Continuous}
We now investigate the absorption spectrum for a monomer coupled to a continuum of Morse oscillators. The system Hamiltonian for the monomer is
\begin{equation}
    H_{S} = \epsilon_{1}\ket{1}\bra{1},
    \label{Eq:SysIndependentSpinBoson}
\end{equation}
with the environment coupling to the excited state via $\ket{1}\bra{1}$. This is equivalent to the independent spin-boson model and allows for the analysis developed in Sec.\eqref{Absorption_Spectra} to determine the absorption spectrum for a continuum of Morse oscillators. We consider the same spectral density as in Fig.~\ref{Fig:Effective_Spectral_Density}(a), with an anharmonicity parameter of $A = 5.1$, and spectral density paramaters $\gamma = 1.0$~ps$^{-1}$, $\lambda = 0.001$~ps$^{-1}$ and $\Omega = 20$~ps$^{-1}$. Fig.~\ref{fig:absorption_spectra_continuous} displays the absorption spectrum for a dimer system coupled to a continuous anharmonic environment at (a) $T = 40$~K and (b) $T = 77$~K. Fig.~\ref{fig:absorption_spectra_continuous}(a) highlights key anharmonic effects present in absorption spectra, particularly the zero-phonon line (centred around $(\omega - \epsilon_{1}) = 0$), which exhibits a finite broadening arising from the non-zero variance of the anharmonic modes. Using a Taylor expansion of the line-shape function Eq.~\eqref{lineshpe_function}, we identify first-order phonon processes within the absorption spectra. Features at higher frequencies correspond to the creation of a phonon, allowing the absorption of a photon by the system through dissipation of energy, whereas at lower frequencies the annihilation of a phonon bridges the gap in the energy required for the photon to be absorbed by the system. Notably, phonon creation (annihilation) occurs at the transition energies of the anharmonic modes, giving rise to a structured absorption spectrum. This structure reflects the non-uniform energy level spacing, which enables a broad range of resonant absorption processes. In addition, we can identify second-order phonon processes, which arise from simultaneous resonant transitions involving two excitations of the anharmonic modes, as seen at $(\omega  - \epsilon_{1}) = 40$~ps$^{-1}$.
We note that phonon processes that occur at larger energy gaps are significantly suppressed due to the decrease in the magnitude of the Morse coupling operator $B_{nm}$. Since the spectral density can be fully determined by using the line-shape function, the absorption spectrum of a monomer coupled to a continuum of anharmonic modes provides direct evidence of the features expected in the effective spectral density. To emphasise the effects of non-zero variance on the absorption spectra, the temperature is increased in Fig.~\ref{fig:absorption_spectra_continuous}(b). As shown in Fig.~\ref{Fig:Effective_Spectral_Density}(a), increasing the temperature of the anharmonic environment enhances the zero-frequency coupling as higher-energy levels of the AMs become occupied, leading to an increase in the variance of the anharmonic modes ensemble. This increase leads to stronger dephasing of the system coherences, as described by Eq.~\eqref{Eq:DecayofCoherences}, producing a broader absorption spectrum. While this broadening can obscure features associated with the underlying spectral density, it provides a quantitative measure of the variance-induced dephasing contribution. This provides a means of probing the underlying structure of the anharmonic mode, as the temperature dependence of the variance contains information about the distribution of the anharmonic energy level structure. From Fig.~\ref{fig:absorption_spectra_continuous}(b), we find the variance through determining the full half-width mask of the absorption spectra around $(\omega -\epsilon_{1}) = 0$ to be $A_{v}^{\text{absorption}} \approx 10.28$~ps$^{-2}$ in close agreement with the exact value of $A_{v}^{\text{exact}} \approx 10.15$~ps$^{-2}$. The small deviation between the exact and computed values can be attributed to higher-order contributions, which introduce additional broadening of the central absorption peak as seen in Fig.~\ref{fig:absorption_spectra_continuous}(b). Therefore, the absorption spectrum provides a direct probe of the energy level structure of the anharmonic modes and enables the extraction of the effective spectral density. It also allows for the quantification of the variance terms that induce static disorder in the system, one of the unique features of anharmonic environments compared to purely harmonic environments.\\

To validate the approach, we employed the ACE algorithm, which allows the environment to be modelled explicitly as a collection of anharmonic modes. Our results demonstrate, however, that the harmonic mapping framework has broader applicability, providing a route for extending numerically exact techniques originally developed for harmonic environments to systems coupled to a continuum of anharmonic modes. 
For approaches that discretise the environment, the zero-frequency contribution can be incorporated explicitly as part of the discretised environment. Alternatively, in methods that trace out environmental degrees of freedom, this contribution can be absorbed into the system Hamiltonian. The latter approach avoids a non-decaying-to-zero correlation function, but increases the computational cost as the system Hilbert space must be enlarged to accommodate the corresponding zero-frequency degree of freedom.

\section{Damped anharmonic modes}
Rather than considering a continuum of anharmonic modes, it can also be advantageous to represent a small number of damped anharmonic modes. This is appropriate when a few discrete damped anharmonic modes strongly interact with the system, while the remaining environment interaction can be captured through an effective dissipative bath. In this section, we develop a framework for describing such damped anharmonic modes, using the damped Morse oscillator as an example.\\

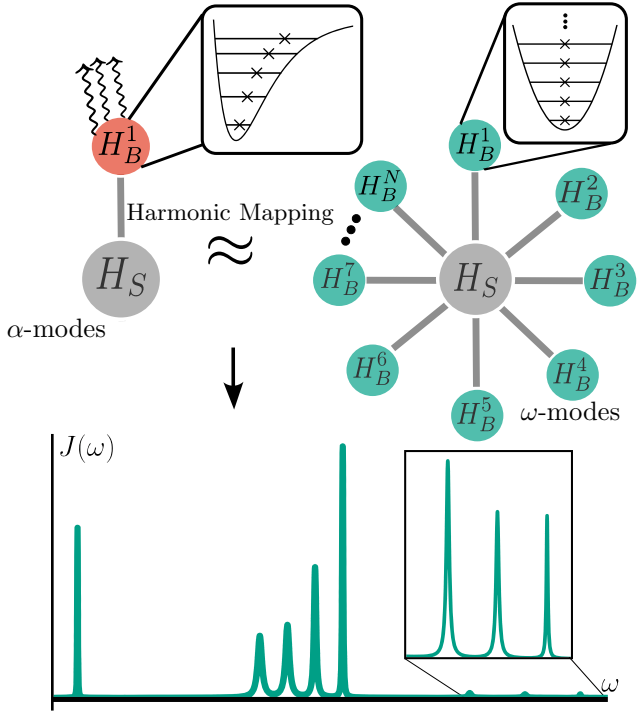
\begin{figure}[!htbp]
    \centering
    \begin{tikzpicture}
    \node[anchor = south west, inner sep =0]
    (fig) at (0,0)
    {\includegraphics[width=1.0\linewidth]{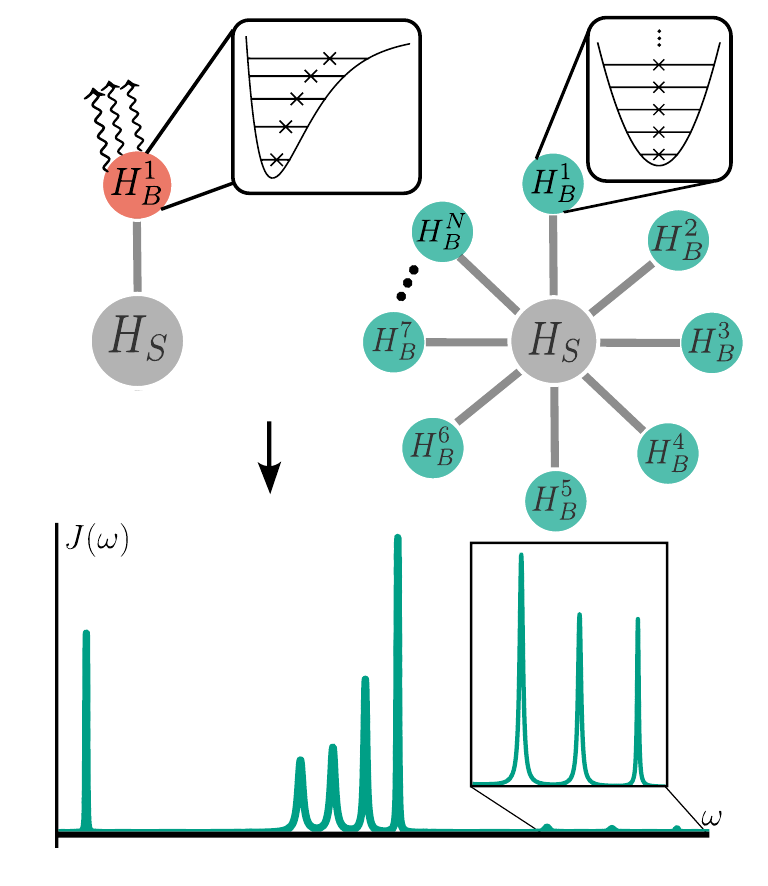}};

    %Text over the figure 
    \node at (0.7,5.5) {$\alpha$-modes};
    \node at (3.0,6.5) {\fontsize{30}{30}\selectfont $\approx$};
    \node at (3.0,7.0) {\small{Harmonic Mapping}};
    \node at (7.5,4.4) {$\omega$-modes};
    \end{tikzpicture}
    \caption{Depiction of the harmonic mapping from a single damped anharmonic mode to an effective harmonic bath. The grey circles represent the system Hamiltonian, which couples to either the red $\alpha$-damped anharmonic mode or the green $\omega$-harmonic modes. Here, $H_B^i$ denotes the bath Hamiltonian of the $i$-th harmonic mode. The damped anharmonic mode comprises a non-Markovian memory core coupled to a Markovian reservoir through which excitations are lost. The insets illustrate the corresponding anharmonic and harmonic potentials. The mapping is constructed by matching the two-time bath correlation function, thereby producing an effective harmonic bath that captures the anharmonic mode up to second-order correlations. Unlike the continuous thermodynamic limit case, where higher-order correlations no longer influence the system dynamics, a finite anharmonic environment retains higher-order correlations, making the mapping approximate. The effective spectral density captures the anharmonicity of the damped mode through shifted peaks associated with its non-equidistant energy-level spacings, while its non-zero variance gives rise to a peak centred around zero frequency. }
    \label{fig:Figure5Draft}
\end{figure}

\subsection {Modelling damped anharmonic modes}
Here, we study damped anharmonic modes that couple to the system of interest. We start by considering a total environment that consists of a non-Markovian core that retains memory effects and a Markovian reservoir in which excitations are lost from the oscillator through dissipation and damping, as depicted in Fig.~\ref{fig:Figure5Draft}. The combination of both environments allows the Markovian reservoir to be traced out explicitly and to be modelled using a Lindblad master equation. For harmonic environments, analogous constructions have been shown to accurately reproduce open-system dynamics and form the basis of pseudomode descriptions ~\cite{DAMPF,NicolaSystematicCoarseGraining}. Within this framework, we adopt a similar effective modelling and investigate the anharmonic effects a damped anharmonic mode can have as an approximate representation of structured anharmonic environments. It should be noted that, much like for an individual harmonic pseudomode~\cite{Cresser01111992}, the individual damped AMs considered here are not thermodynamically complete, as the dissipator acts only on the AM and not the joint system-AM Hilbert space. However, like harmonic pseudomodes it can provide an effective framework that captures non-Markovian features that can appear due to the anharmonicity of a damped anharmonic mode. For an arbitrary interaction Hamiltonian, we have

\begin{equation}
    H_{I} = \sum_{i=1}^{N}g_{i}{S_{i} \otimes B_{i}} , 
    \label{eq:Interaction_Environment_Damped}
\end{equation}
where $S_{i}$ is the system operator that couples to the $i^\text{th}$ anharmonic mode with bath coupling operator $B_{i}$ and coupling strength $g_{i}$. Each anharmonic mode is then coupled to independent Markovian thermal baths which,  after tracing out, is modelled via the Lindblad equations, 

\begin{gather}
    \sum_{n>m}^{N_{\alpha} -1}\mathcal{D}_{nm\uparrow}\rho
    =
    \begin{aligned}[t]
    \sum_{n>m}^{N_{\alpha} -1}\gamma_{\uparrow}(\alpha_{nm})\Bigl(
    &B_{\alpha,nm}\rho B_{\alpha,nm}^{\dagger} \\
    &-\frac{1}{2}\{B_{\alpha,nm}^{\dagger}B_{\alpha,nm},\rho\}
    \Bigr),
    \end{aligned}
    \\[1ex]
    \sum_{n<m}^{N_{\alpha} -1}\mathcal{D}_{nm\downarrow}\rho
    =
    \begin{aligned}[t]
    \sum_{n>m}^{N_{\alpha} -1}\gamma_{\downarrow}(\alpha_{nm})\Bigl(
    &B_{\alpha,nm}\rho B_{\alpha,nm}^{\dagger} \\
    &-\frac{1}{2}\{B_{\alpha,nm}^{\dagger}B_{\alpha,nm},\rho\}
    \Bigr),
    \end{aligned}
\label{Eq:Lindblad_Terms}
\end{gather}
where $B_{\alpha,nm}$ are the matrix elements of the coupling operator for a single damped anharmonic mode $B_{\alpha,nm} = B_{nm}(\alpha)\ket{n}\bra{m}$. We choose $\gamma_{\uparrow}(E_{nm}(\alpha)) = \gamma n_{th}(E_{nm}(\alpha))$ and $\gamma_{\downarrow, nm} = \gamma (n_{th}(|E_{\alpha, nm}|) + 1)$ to ensure detailed balance and $n_{th}(E_{\alpha, nm})$ is the Bose-Einstein occupation of the mode gap $E_{nm}(\alpha)$. These Lindblad terms are derived from the Bloch-Redfield equation for the single damped anharmonic mode coupled to the thermal Markovian structureless harmonic bath. See Appendix~\ref{App:DampME} for details on the Lindblad form.
\begin{figure*}
    \centering
    \includegraphics[width=\linewidth]{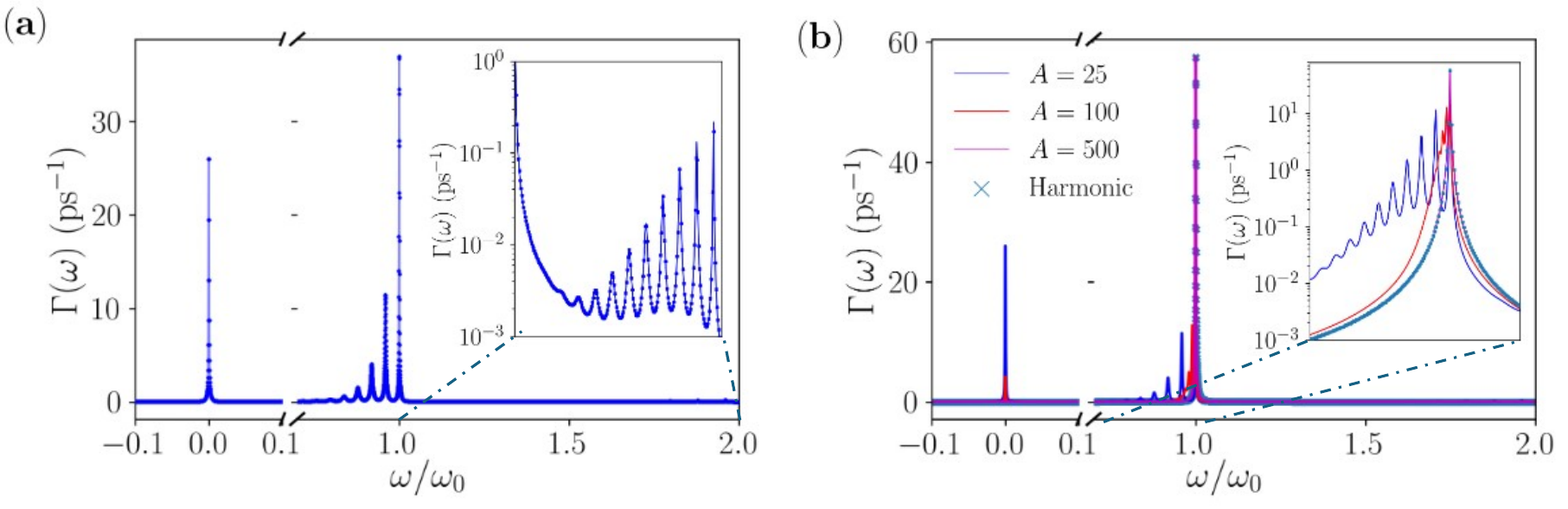}
    \caption{(a) Calculation of the real part of the Fourier transformed two-time correlation function for a Morse oscillator calculated using the approximate Laplace transform (dot) and numerically (full). $\alpha = 20$~ps$^{-1}$ $\gamma=0.0005~\alpha$, $\beta = 1/ \alpha$, $A = 25.1$. The insert shows the blue-shifted peaks on a logarithmic scale. (b) The spectral function of a single damped Morse oscillator environment with varying anharmonicity parameters $A$. The underlying frequency of the mode is $\alpha = 20$~ps$^{-1}$ with a decay rate of $\gamma = 0.01$~ps$^{-1}$ at $\beta = 1/(k_b\alpha)$. The insert demonstrates the enhanced anharmonic effects around the transition frequency of the oscillator for varying anharmonicity parameter $A$.}
    \label{fig:DampedMorseOscillatorSpectrum}
\end{figure*}
\subsection{Effective spectral density of damped anharmonic modes}

As in the continuous case, it is relevant to calculate the two-time bath correlation function for a damped anharmonic mode to understand how anharmonic environment effects manifest, leading to unique non-Markovian system dynamics. Once the two-time correlation function has been calculated, it can be used to generate an effective spectral density, as outlined in Section I, enabling the single damped anharmonic mode to be modelled as a continuum of harmonic oscillators.  We note, however, that the mapping is expected to be incomplete for the single damped anharmonic modes, as the environment is no longer in the thermodynamic limit ($N\rightarrow \infty$); hence, higher-order correlation effects will contribute to the system's dynamics. We begin with the damped anharmonic mode-free Hamiltonian,
\begin{equation}
    H_B(\alpha) = \sum_{n=0}^{N_{\alpha}-1}E_n(\alpha)\ket{n_{\alpha}}\bra{n_{\alpha}} ,
\end{equation}
where $E_n(\alpha)$ is the energy of the $n$th level of the $\alpha$ anharmonic mode. The anharmonic mode then undergoes dephasing, dissipation, and thermal fluctuations through the Lindblad dissipators from Eq.~\eqref{Eq:Lindblad_Terms}. The Lindblad master equation for a single damped anharmonic mode is of the form, 
\begin{equation}
    \begin{split}
    \frac{\partial \rho}{\partial t}
    =
    -i[H_B(\alpha),\rho]
    + \sum_{n<m}^{N_{\alpha}-1}\mathcal{D}_{\downarrow}\rho
    + \sum_{n>m}^{N_{\alpha}-1}\mathcal{D}_{\uparrow}\rho .
    \end{split}
\end{equation}
 In order to calculate the effective spectral density for this system, we require the two-time bath correlation function. Due to the presence of Lindblad operators, we calculate the damped anharmonic mode correlation function in the Heisenberg frame via
\begin{align} 
    C(t) &= \langle B_\alpha(t)B_\alpha(0)\rangle\nonumber\\&=\Tr{B_{\alpha}\rho^{Q}(t)},
\label{Eq:Linblad_Correlation_Solver}
\end{align}
where $B_{\alpha}$ is the coupling operator for a single damped anharmonic mode and $\rho^{Q}(t)$ is a quasi-density operator with initial state given by
\begin{equation}
    \rho^{Q}(0) = B_{\alpha}\rho_{\beta} = \sum_{n,m=0}^{N_{\alpha}-1} B_{\alpha, nm} p_{m} \ket{n}\bra{m}
\end{equation}
with $p_{m} = \frac{e^{-\beta E_m(\alpha)}}{Z}$ and $Z = \sum_{m = 0}^{N_{\alpha} -1}e^{-\beta \alpha E_m(\alpha)}$. The Lindblad master equation above can be shown to preserve the structure of the quasi-density matrix such that its time evolution is of the form,
\begin{equation}
    \rho^{Q}(t) = \sum_{n,m=0}^{N_{\alpha}-1}R_{nm}(t)\ket{n}\bra{m},
\end{equation}
thus the two-time bath correlation function can be evaluated as 
\begin{equation}
    C(t) = \sum_{n,m=0}^{N_{\alpha}-1}R_{nm}(t)B_{nm}
\end{equation}
with $R_{nm}(0) = B_{\alpha, nm}p_{m}$ and $R_{nm}(t)$ remains to be calculated. Through application of the Lindblad master equation, we can derive a set of differential equations for the elements of $R_{nm}$, see Appendix~\ref{App:DampSpecDens} for full details, which then allows for the calculation of the rate function via 
\begin{equation}
\begin{split}
\Gamma(\nu) &= \Re{\int dt\, e^{-i\nu t} C(t)} \\
&= \Re{\sum_{n,m=0}^{N_{\alpha}-1} \int dt\, B_{\alpha,nm}
R_{nm}(t) e^{-i\nu t}}.
\end{split}
\end{equation}
Separating the rate function into diagonal $(n = m)$ and off-diagonal components $(n \neq m)$ leads to the following rate equations, 

\begin{equation}
    \Re(\Gamma(\nu))
    \approx
    \sum_{n\neq m = 0}^{N_{\alpha}-1}
    \frac{|B_{\alpha,nm}|^{2}p_n\gamma_{nm}}
    {(\nu+E_{\alpha, nm})^{2}+\gamma_{nm}^{2}}
    +
    \sum_{i=0}^{}
    \frac{X_i\lambda_i}
    {\nu^{2}+\lambda_i^{2}} ,
\label{Eq:TotalSpectralFunction}
\end{equation}
where $\gamma_{nm}$ captures all absorption and decay transition rates associated with the energy levels $n$ and $m$, and the final term within Eq.~\eqref{Eq:TotalSpectralFunction} is a contribution from the diagonal elements of the bath coupling operator, which acts as a low-frequency Lorentzian dephasing term, similar to the divergent term found in the continuous case. The rate equations directly relate to the two-time frequency-domain correlation function for a damped anharmonic mode and can be mapped onto a thermalised harmonic environment through Eq.~\eqref{Eq:thermalisedspectraldensity}\\
\begin{equation}
    J^\text{th}_\text{eff}(\omega, \beta) = \frac{1}{\pi}\Re(\Gamma
(\omega)).
\label{damped_harmonic_spectral_density}
\end{equation}
As in the continuous anharmonic modes discussed previously, the steady state of the two-time bath correlation function does not decay to zero due to the non-zero diagonal components of the coupling operator. Specifically, the steady state corresponds to the Lorentzian centred around $\omega = 0$ when $\lambda_{i}=0$; therefore, to mitigate this divergence within our spectral function, we again displace our coupling operator by the thermal average. Unlike the continuous case, this amounts to zeroing the steady-state value of the bath correlation function as,
\begin{align}
    \hat{C}(t) &= \langle (B_\alpha(t)-\langle B_\alpha\rangle)(B_\alpha(0)-\langle B_\alpha\rangle)\rangle \\& = \langle B_{\alpha}(t) B_{\alpha}(0)\rangle - \langle B_{\alpha} \rangle ^{2}
\end{align}
which is equivalent to, 
\begin{align}
   C(t) - C(\infty)&= \langle B_{\alpha}(t)B_{\alpha}(0)\rangle 
   - \lim_{t \rightarrow \infty} \langle B_{\alpha}(t)B_{\alpha}(0)\rangle \nonumber\\
   &= \langle B_{\alpha}(t)B_{\alpha}(0)\rangle 
   - \lim_{t \rightarrow \infty} \langle B_{\alpha}(t)\rangle \langle B_{\alpha}(0)\rangle \nonumber\\
   &= \langle B_{\alpha}(t)B_{\alpha}(0)\rangle 
   -\langle B_{\alpha}\rangle^2,
\end{align} 
where the large time separation leading to a product decomposition manifests due to the relaxation of the anharmonic mode. This removes the divergence from the effective spectral density that persisted in the continuous case leaving only a continuous spectrum.

\begin{figure*}[ht]
    \centering
    \includegraphics[width=\linewidth]{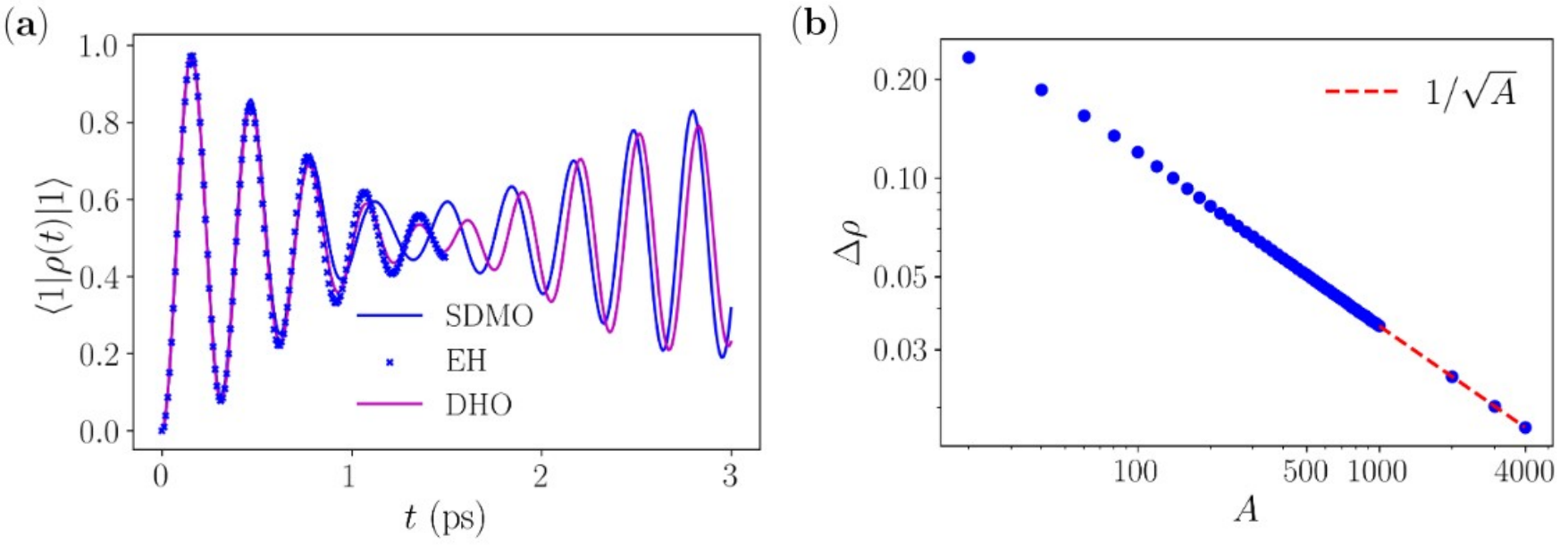}

    \caption{\label{fig:DampedMorseOscillatorDynamics} (a) Dynamics of a dimer system coupled to a single damped anharmonic Morse oscillator (SDMO) with anharmonicity parameter $A = 25.1$, the dynamics from an effective harmonic environment (EH) where the effective spectral density is produced from a single damped anharmonic Morse oscillator with the dynamics from a damped harmonic oscillator (DHO). (b) Absolute difference between damped anharmonic with varying anharmonicity parameters and damped harmonic oscillator $\Delta \rho =| \bra{1}\rho_{\text{anharmonic}}\ket{1} - \bra{1}\rho_{\text{harmonic}}\ket{1}|$ highlighting that even deep potential wells of damped anharmonic modes can have substantial anharmonicity within them. The parameters of the damped Morse and harmonic oscillator is $\alpha, \omega =20$~ps$^{-1}$, $\gamma = 0.01$~ ps$^{-1}$, $\beta = \hbar /(k_{b} \omega_{0})$, $g = 2.5$~ps$^{-1}$, with system parameters $\epsilon_{1} = \epsilon_{2} = 0 $ and $2\Delta = 20$~ps$^{-1}$.}  
\end{figure*}

\subsection{Effective Spectral density for a damped Morse oscillator}

Using the framework described above, we can analytically evaluate the rate functions for a single damped Morse oscillator by using Eq.~\eqref{Eq:TotalSpectralFunction}. We can then numerically validate this by computing the rate functions by solving the Lindblad master equation and evaluating Eq.~\eqref{Eq:Linblad_Correlation_Solver}, where taking the real part of the Fourier transform produces the rate functions. We find that the secular approximation solution yields good agreement with a numerical approach, as shown in Fig.~\ref{fig:DampedMorseOscillatorSpectrum}(a). As for the continuous case, we highlight that the non-equidistant level structure of the Morse oscillator and thermal occupations of higher levels reveal additional red- and blue-shifted structure within the spectral density, as shown in Fig.~\ref{fig:DampedMorseOscillatorSpectrum}(b). As the anharmonicity parameter $A$ increases, the spectrum approaches the harmonic limit as the energy level spacings become equal. Concurrently, the zero-frequency contribution diminishes, reflecting the vanishing diagonal matrix elements of the Morse oscillator in the large-$A$ limit ($x_{nn} \to 0$). However, we find that to achieve the full harmonic approximation, the potential depth of the Morse potential must be extremely deep. This can be seen from taking the asymptotic limit of the coupling operator; the off-diagonal elements of the operator become  
\begin{equation}
    {\lim_{A \rightarrow \infty}}x_{nm} = \frac{(-1)^{(m-n+1)}}{(m-n)} (2A)^\frac{n-m}{2}\sqrt{\frac{m!}{n!}}.
\end{equation}
The decay of the anharmonic contributions to the off-diagonal elements of the coupling operator is controlled by the energy level spacings. In particular, transitions between widely separated levels decay significantly faster than those between neighbouring levels, which scale as $\propto 1/\sqrt{2A}$ since $m > n$. In addition, we find the same decay for the off-diagonal terms.

\subsection{Dynamics of a dimer coupled to a single damped Morse oscillator}
Using Eq.~\eqref{damped_harmonic_spectral_density}, we can then enable a direct comparison between the dynamics of a single damped Morse oscillator and the mapping onto a continuous bosonic bath of harmonic oscillators. The latter captures only the behaviour of the two-time correlations; thus it assumes higher-order correlations of the damped Morse oscillator have decayed to zero or can be expressed in terms of two-time correlations. To highlight the differences between the damped Morse mode and the mapping to a continuous bosonic environment, we also include the dynamics of a damped harmonic mode using equations Eq.~\eqref{eq:Interaction_Environment_Damped} and Eq.~\eqref{Eq:Lindblad_Terms} with the coupling operator being the harmonic coupling operator. The dynamics for each representation of the environment are again obtained using the ACE formalism. Within ACE, we incorporate the effects of the Markovian environment by including the appropriate Lindblad terms in the environment's Liouville operator. The system of interest is again a homodimer in the single excitation manifold given by Eq.~\eqref{Eq.Molecular_Dimer} with $\epsilon_{1} = \epsilon_{2} = 0$ and $2\Delta = 20$~ps$^{-1}$.\\

\begin{figure}
    \centering
    \includegraphics[width=\linewidth]{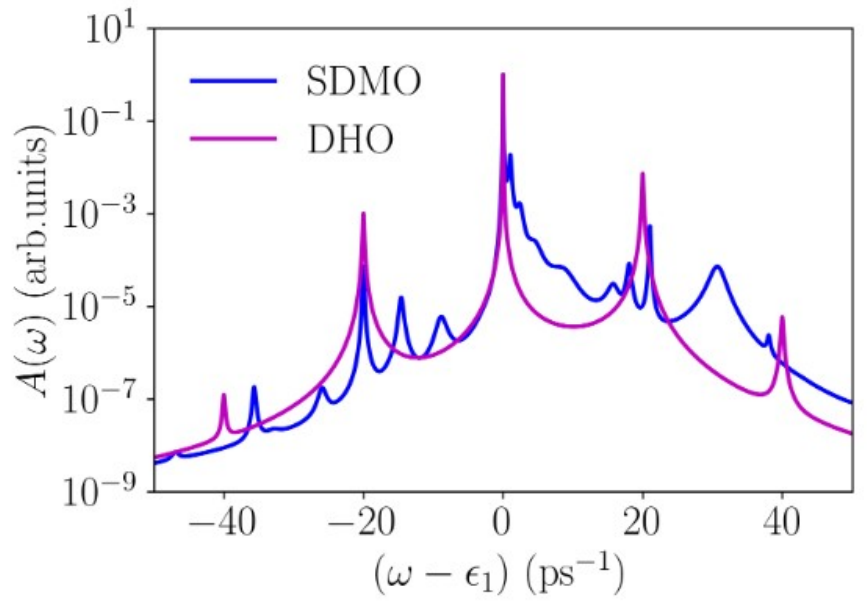}
    \caption{Absorption spectra for a single damped Morse oscillator (SDMO) environment with anharmonicity $A = 5.1$ and a single damped harmonic oscillator (DHO). The underlying frequency of the environment mode is $\alpha_{0} = 20$~ps$^{-1}$, $g = 1.0$~ps$^{-1}$ with a overall decay rate of $\gamma_{0} = 0.01$~ps$^{-1}$ at $T = 77$~K with system parameters $\epsilon_{1} = 20$~ps$^{-1}$ and $\Delta_{i} = 0$.}
    \label{fig:AbsorptionSpectraDampedOscillators}
\end{figure}

\begin{figure*}[ht]
    \centering
    \includegraphics[width=0.8\linewidth]{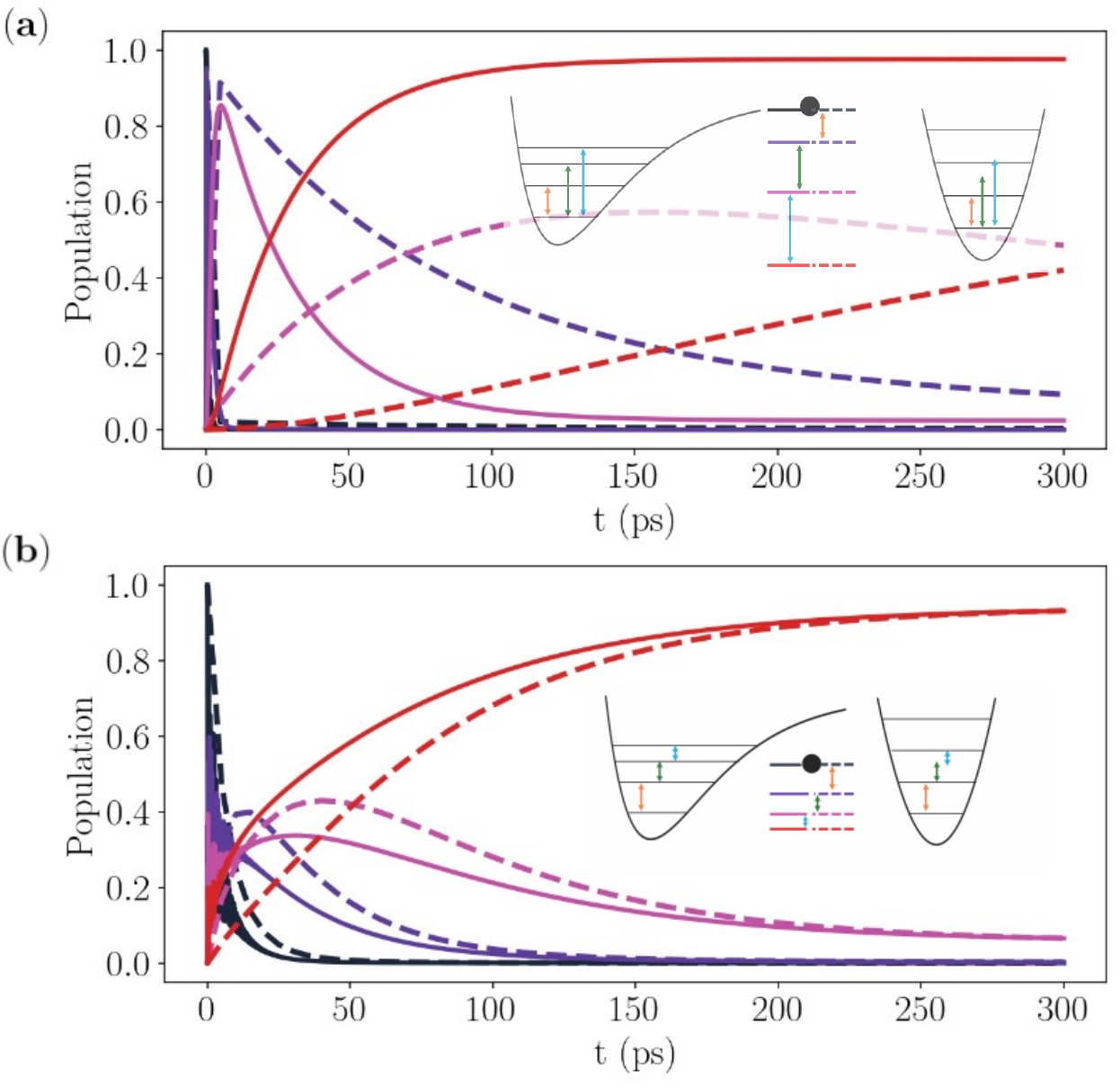}
    \caption{Population dynamics of a tetramer in the single-excitation manifold, initially prepared in the highest-energy system state, coupled to either a single damped Morse oscillator (solid lines) with a number of bound states of $A=10.1$ or a single damped harmonic oscillator (dotted lines). The oscillators have a ground-to-first-excited-state transition frequency of $\alpha, \omega =151.93$~ps$^{-1}$, decay rates of (a) $\gamma = 3.04$~ps$^{-1}$, (b) $\gamma = 0.30$~ps$^{-1}$ and coupling strengths of (a) $g = 4.55$~ps$^{-1}$, (b) $g = 10.64$~ps$^{-1}$ and are at $T=300$~K. (a) Population dynamics when the system energy gaps are resonant with the blue-shifted transition frequencies of the anharmonic mode. The inset highlights the resonant system–environment transitions of the anharmonic oscillator and corresponding off-resonant transitions of the harmonic oscillator, with system parameters $\epsilon_0=0$, $\epsilon_1=E_{\alpha,01}$, $\epsilon_2=E_{\alpha,02}$, and $\epsilon_3=E_{\alpha,03}$. (b) Populations when the system energy gaps are resonant with the red-shifted transition frequencies of the anharmonic mode. The inset highlights the corresponding resonances, with $\epsilon_0=0$, $\epsilon_1=E_{\alpha,01}$, $\epsilon_2=E_{\alpha,12}$, $\epsilon_3=E_{\alpha,23}$, and $\Delta_i=0$.}
    \label{fig:4_level_dynamics}
\end{figure*}

In Fig.~\ref{fig:DampedMorseOscillatorDynamics}(a), we compare the dynamics obtained from the three representations of the environment. At earlier times, all three scenarios were in good agreement; however, clear deviations emerged at later times. These deviations are expected, particularly between the damped anharmonic and damped harmonic oscillator models. As in the continuous case, the effective coupling of the resonant peak is reduced relative to the harmonic approximation, resulting in a slower decay of the system population dynamics. 
Since the system is coupled to a single non-Markovian environment, coherent excitations are exchanged between the system and the environment. If the system-environment coupling exceeds the oscillator damping rate, coherent excitation transfer dominates over dissipation of the oscillator into the Markovian environment, resulting in beating dynamics where excitations from the environment transfer back into the system. Over long timescales, these oscillations are damped as the environment loses excitations to the surrounding Markovian environment, ultimately driving the system towards a steady state. Comparing the damped anharmonic mode with the mapping framework, we observe that at later times the dynamics begin to deviate. This behaviour is expected, as the mapping assumes the environment is Gaussian and does not capture higher-order correlations that affect the system's dynamics. For the mapping framework, we only propagate the dynamics up to $t_{final} = 1.5$~ps as convergence to later times becomes numerically difficult. Fig.~\ref{fig:DampedMorseOscillatorDynamics}(b) shows that as the anharmonicity parameter of the damped Morse oscillator increases (corresponding to a deeper well), the deviation between the harmonic and anharmonic dynamics over the full propagation time decreases $\propto 1 / \sqrt{A}$. For the parameters considered, an error of approximately $10\%$ between the harmonic and anharmonic dynamics is obtained at $A=199.1$. Reducing the error to $2\%$ requires a substantially larger anharmonicity parameter of $A=2999.1$. Overall, this further highlights the need for accurate simulations of anharmonic modes, as even small anharmonic effects take place in deep wells.

\subsection{Absorption spectrum for a monomer coupled to a single damped Morse oscillator}

We now investigate the absorption spectrum for a single damped Morse oscillator coupled to a monomer system to analyse the effects of anharmonicity within damped anharmonic modes. The system Hamiltonian is chosen to be the same as Eq.~\eqref{Eq:SysIndependentSpinBoson} such that it is in the independent spin-boson model with $\epsilon_{1} = 20$~ps$^{-1}$ and $\Delta = 0$ to allow for environmental contributions to be easier to distinguish. The damped anharmonic oscillator parameters are taken to be similar to those in Fig.~\ref{fig:DampedMorseOscillatorDynamics}(a), with the anharmonicity and temperature modified to $A = 5.1$ and $T = 77$~K, respectively.

Fig.~\ref{fig:AbsorptionSpectraDampedOscillators} demonstrates the absorption spectra for a single damped Morse oscillator and that of a single damped harmonic oscillator with parameters specified in Fig.~\ref{fig:AbsorptionSpectraDampedOscillators}. In both cases, the zero-phonon absorption line at $\omega-\epsilon_{1} = 0$ corresponds to a resonant photon absorption by the system. For the damped harmonic oscillator, the absorption spectrum exhibits the same behaviour as discussed in Sec.~\ref {Absorption_Continuous}, where photon absorption is accompanied by the creation (annihilation) of phonons at transition frequencies within the oscillator. In contrast, the absorption spectrum for the damped Morse oscillator reveals notable differences compared to the damped harmonic oscillator. In this case, it is more appropriate to consider the joint eigenstructure of the system and the Morse oscillator, as diagonalization of the system–environment interaction leads to hybridisation of their respective energy levels. We find that the absorption lines correspond to transitions between eigenenergies of the combined system and Morse oscillator, leading to many additional, distinct absorption features across the frequency range corresponding to the absorption of a photon in the hybridised system. The number of distinct spectral features depends sensitively on the anharmonicity of the damped Morse oscillator, as the non-uniform energy level spacings allow photon absorption across a range of the transition energies of the hybridised system. As the anharmonicity parameter increases (larger potential depth), these additional distinct absorption peaks become less prominent as the energy gaps between the levels in the anharmonic oscillator become harmonic, until in the $A \rightarrow \infty$ limit we recover the harmonic absorption spectrum. This is readily understood by noting that the polaron transformation diagonalises the system-environment interaction for the harmonic oscillator case, leading only to an energy shift of the monomer by the reorganisation energy; the harmonic oscillator then acts like a free oscillator, leading to the separability of the energy levels, unlike in the anharmonic case.

\subsection{Energy transport within a four-level system}
Quantum transport is a highly relevant feature of many large quantum networks, such as excitation transfer in light-harvesting systems, and understanding its behaviour is essential for designing efficient quantum devices ~\cite{AChin_ENAQT, RouseGaugerLovett2019, BurgessGauger2026}. In particular, both theoretical ~\cite{PlenioHuelga2008, ENAQT, Zerah-HarushDubi2018} and experimental ~\cite{Sowa2017, Biggerstaff2016} studies have investigated environment-assisted quantum transport (ENAQT), in which interactions with the environment can enhance transport efficiency. In particular, within ENAQT, the energy landscape critically influences transport efficiency. Therefore, to highlight different, distinct effects of anharmonicity relative to the harmonic approximation, we consider a four-site model in the single-excitation manifold and investigate how a damped anharmonic mode, as opposed to a damped harmonic oscillator, affects the transport efficiency. The Hamiltonian for the four-site model in the single-excitation manifold is described by  
\begin{equation}
    H_{S} = \sum_{i = 0}^{3} \epsilon_{i}\ket{i}\bra{i} + \sum_{i < j}^{3}\Delta_{i} (\ket{i}\bra{j}) + h.c., 
\end{equation}
with the coupling to the environment modified to induce transitions between energy levels through $S = \ket{0}\bra{1} + \ket{1}\bra{2} + \ket{2}\bra{3} + h.c.$. We choose $\Delta_{i} = 0$, so that transitions between sites are solely through the interaction with the environment. We choose the site energies $\epsilon_{i}$ such that, arranged in ascending order, the transitions are resonant with the red- or blue-shifted transitions in the damped anharmonic oscillator. \\

Fig.~\ref{fig:4_level_dynamics}(a-b) shows population transfer from the excited to ground state for a system coupled to either a damped anharmonic mode (solid) or a damped harmonic mode (dashed). The inserts show the energy level structure of the four-level system, where for (a) the site energy differences are resonant with the blue-shifted transitions of the damped anharmonic mode while in (b) they are resonant with the red-shifted transitions. Included in the insert is the energy level structure of the damped harmonic oscillator where the system couples to off-resonant transitions of the damped harmonic mode. The corresponding system parameters are indicated in the figure. Fig.~\ref{fig:4_level_dynamics} demonstrates that coupling to an anharmonic environment leads to a significant increase in population transfer from the excited to the ground state compared to the harmonic case. This enhancement arises from the non-uniform level spacing of the anharmonic oscillator, which enables the system's transition energies to be resonant with multiple environmental transitions. For Fig.~\ref{fig:4_level_dynamics}(a), the decay rate of the damped oscillator is comparable to the rate of the system-environment coupling strength. Consequently, environmental excitations relax to the ground state on a similar timescale to that associated with their population, preventing the accumulation of excitations to higher excited levels in the anharmonic oscillator. This leads to efficient energy transfer dynamics in the system: initially, the system can transfer along the ladder through a resonant transition with the ground-to-first excited state of the oscillator. Owing to the rapid decay of the anharmonic oscillator, this excitation quickly returns to the ground state, enabling subsequent resonant transitions with higher excited-state transitions within the system.  This sequential cascade facilitates efficient population transfer from the excited state to the ground state, where, for the anharmonic case, near unity transfer occurs at $t = 100$~ps compared to only $0.1\%$ transfer for the harmonic case at $t = 100$~ps  The harmonic environment provides only a single resonant condition with the harmonic oscillator; therefore, the rapid decay of excitations within the environment no longer leads to multiple resonant pathways for the system to utilise, and the enhanced transport is suppressed. \\

We further examine the transport dynamics when the system Hamiltonian is tuned to be resonant with red-shifted transitions in the anharmonic oscillator shown in Fig.~\ref{fig:4_level_dynamics}(b). In this regime, the system-environment coupling exceeds the decay rate of the anharmonic oscillator, enabling population of higher energy levels in the anharmonic oscillator. Resonant transport in the system is facilitated by upward transitions within the anharmonic ladder. Although transport to the ground state is enhanced relative to the harmonic oscillator, the enhancement is less pronounced than in the blue-shifted configuration. This reduction arises because the relevant system transition frequencies are resonant with adjacent level spacings, requiring the system-environment interaction to promote excitations to higher levels before relaxation takes effect. As a result, energy transfer is less efficient than in the blue-shifted case, where rapid decay directly assists the transport process, whereas here it instead hinders it.
\begin{figure}
    \centering
    \includegraphics[width=\linewidth]{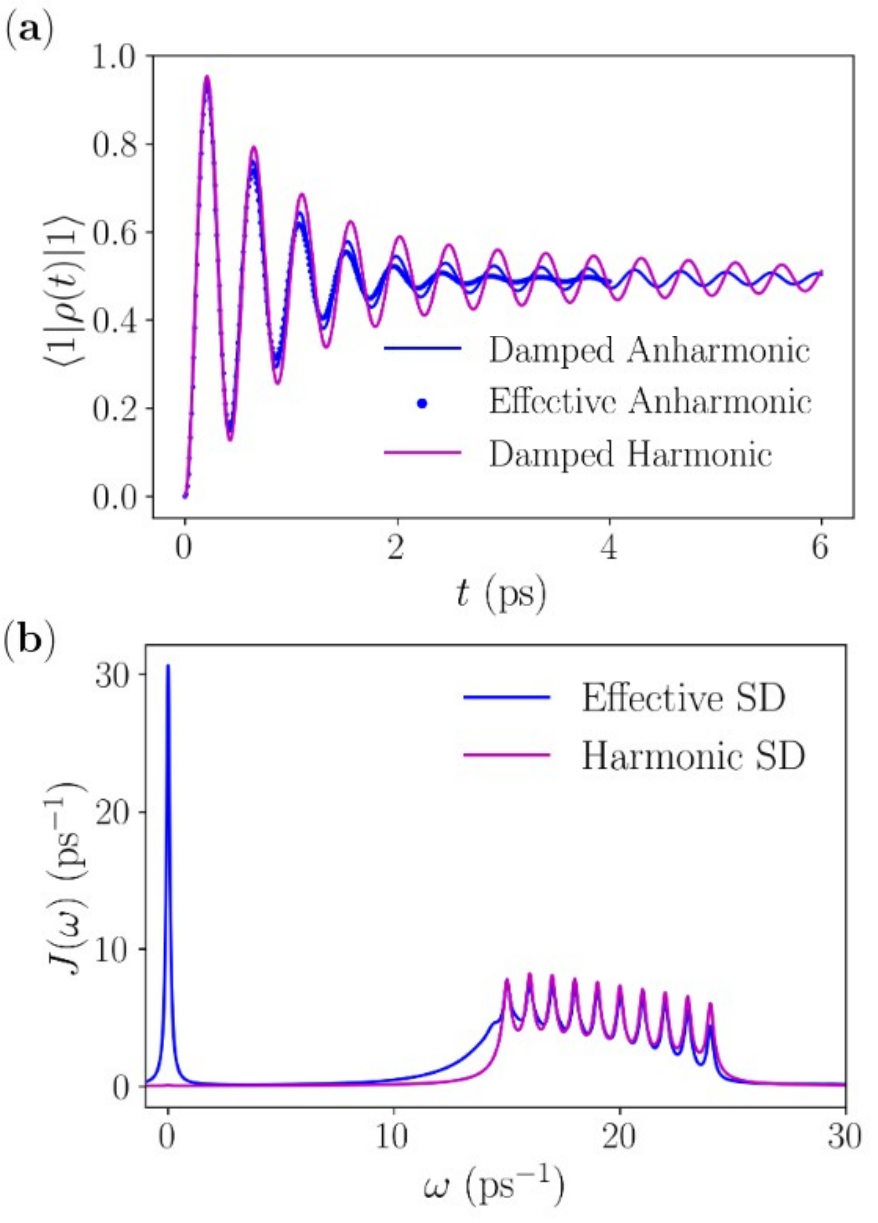}
    \caption{Dynamics of a dimer coupled to ten damped Morse oscillators ($A = 25.1$) compared with ten equivalent damped harmonic oscillators. Oscillator ground to excited state transition frequencies are $\alpha_i = {15, 16, \dots, 24},~\mathrm{ps}^{-1}$ with decay rates $\gamma_i = 0.005~\alpha_i$, temperature $T = 150~\mathrm{K}$, and system--bath coupling $g = 1.0$~ps$^{-1}$. The dimer parameters are $\epsilon_1 = \epsilon_2 = 0$ and $2\Delta = 15\mathrm{ps}^{-1}$. The insert shows the spectral functions for both anharmonic and harmonic environments.}
    \label{fig:ManyDampedAnharmonicOscillator}
\end{figure}

\subsection{Many damped anharmonic modes}

When modelling an environment, a single damped anharmonic mode may be insufficient to accurately account for the complexity of the entire environment. It may be more appropriate to consider a collection of many damped modes, with each mode representing a distinct environmental degree of freedom. For example, in biomolecular systems, many vibrational features contribute to the environment, and each feature may require its own damped mode to accurately describe its influence on the system dynamics ~\cite{NicolaSystematicCoarseGraining}. Therefore, we compare the dynamics of many damped anharmonic and harmonic oscillators and map the resulting spectral function of the anharmonic modes onto a continuous harmonic bath using the approach outlined above. We revert to a dimer in the single excitation manifold as our system of interest Eq.~\eqref{Eq.Molecular_Dimer} with parameters $\epsilon_{0} = \epsilon_{1} = 0$ and $2\Delta = 15$~ps$^{-1}$. Fig.~\ref{fig:ManyDampedAnharmonicOscillator}(a) demonstrates the dynamics of the dimer system coupled to an environment consisting of ten damped anharmonic or harmonic oscillators depicted within Fig.~\ref{fig:ManyDampedAnharmonicOscillator}(b), with frequencies and damping rates specified in the figure. It also includes the dynamics obtained from the mapped continuous bath constructed from the anharmonic rate function Eq.~\eqref{Eq:TotalSpectralFunction}. The system's parameters are chosen such that the transition energy is resonant with the lowest-frequency damped oscillator, and the population dynamics of the monomer are given over time.\\

We observe, for the system parameters chosen, that the anharmonic environment leads to a faster decay of the system's dynamics compared to the harmonic case. This is counter to the intuition of the harmonic approach, where, typically, the coupling strength of the resonant oscillator is reduced compared to the harmonic case. However, here the coupling strength is not reduced to the same extent as shown in Fig.~\ref{fig:ManyDampedAnharmonicOscillator}(b), indicating that the observed differences in the dynamics arise from intrinsic anharmonic effects, including higher-order correlations. Notably, the mapped dynamics exhibit stronger damping compared to the exact anharmonic case, resulting in a faster decay to the steady state, reflecting the neglect of higher-order correlations present in the anharmonic oscillator. This further underpins the importance of accurately accounting for anharmonic environments, both in the continuous limit and for a discrete number of anharmonic modes. In particular, increasing the number of oscillators does not necessarily ensure that the mapping approach accurately captures the dynamics. This limitation arises because the mapping framework requires the system–environment coupling to scale as $g/\sqrt{N}$, a condition that is not generally satisfied for a finite set of discrete oscillators, thereby underscoring the need to model both continuous and finite anharmonic environments.

\section{Summary \& Discussion }
In conclusion, we have developed a framework for the numerically exact treatment of anharmonic environments based on a matching criterion for a continuum of anharmonic modes. When higher-order correlations vanish, this criterion enables the construction of a general effective harmonic spectral density. The resulting spectral density contains multiple contributions derived from the bare spectral density, shifted according to the non-equidistant energy gaps of the anharmonic modes, and can therefore exhibit substantially richer structure than its harmonic counterpart. We additionally identify a unique zero-frequency contribution arising from the non-zero variance of the anharmonic modes' coupling operator. This contribution acts effectively as static disorder on the system and can be incorporated efficiently into numerically exact approaches, either through the explicit inclusion of a zero-frequency mode in the environmental discretization or by absorbing its effect into the system Hamiltonian. Together, these results provide a general route for treating anharmonic environments and extend the applicability of methods originally developed for harmonic baths.

To demonstrate the framework, we employed ACE, which allows both the effective harmonic environment and the underlying anharmonic environment to be treated explicitly. We showed that anharmonicity produces distinct non-Markovian features in the reduced system dynamics that are absent in the corresponding harmonic environment. Analysis of the absorption spectrum further reveals additional broadening around zero frequency associated with the zero-frequency contribution, providing a direct spectroscopic signature of this effect.

In parallel, we developed a framework for studying finite collections of damped anharmonic modes beyond the continuum limit. This highlights the importance of considering not only continuous anharmonic environments but also finite sets of anharmonic modes. As in the continuum case, anharmonicity introduces additional spectral features arising from the unequal energy-level spacings of the modes, thereby modifying the effective system--bath couplings and influencing decoherence and energy transfer in ways that cannot be captured by purely harmonic models.

Overall, this work advances our understanding of how anharmonic environments, whether continuous or composed of finite collections of damped modes, influence open-system dynamics beyond the harmonic approximation. Several avenues for future work follow naturally from these results. The continuous framework could be used to investigate open quantum systems such as the Bose--Hubbard model in solid-state settings, or more complex environments in which the bath degrees of freedom interact with one another, as encountered in optomechanical and spin systems. Within the finite-mode framework, it would also be valuable to investigate energy transport in systems comprising multiple monomers coupled to their own anharmonic vibrational environments. Such models are directly relevant to biomolecular aggregates, including the Fenna--Matthews--Olson (FMO) complex, a prototypical system for studying exciton energy transfer and quantum effects in biology, where anharmonic vibrational effects may play an important role~\cite{Engel2007,Scholes2011}.

\section{Acknowledgments}

KD, EMG and AB acknowledge the Leverhulme Trust (Grant No.~RPG-2022-335), the Volkswagen Foundation (Grant No.~0200195) and EPSRC (Grant No. EP/W524669/1). BWL acknowledges funding from EPSRC (Grant No.~UKRI3778).
 
\newpage

\section*{Appendix}
\appendix

\section{Derivation of the influence functional for an anharmonic environment in the thermodynamic limit}\label{App:InfFunc}

Consider an open quantum system model with the following Hamiltonian,

\begin{equation}
    H = H_{S} + H_{B} + H_{I},
\end{equation}
where $H_{S}$ is the system of interest, $H_{B}$ the environment Hamiltonian for a single mode defined in Eq.~\eqref{Eq:Environment_Hamiltonian} and $H_{I} = S \otimes gB$ with $B$ being the bath coupling operator.
The evolution of the density matrix $\rho(t)$, where $\rho(t)$ comprises both the system and environment states, can be calculated as follows. We begin with the usual von Neumann equation
\begin{equation}
    \frac{d \rho(t)}{dt} = -i[H, \rho(t)],
    \label{Time_Evolution}
\end{equation}
where $\hbar =1$ for this derivation. Moving into the interaction picture with respect to both the system and environment Hamiltonian allows us to reformulate Eq.~\eqref{Time_Evolution} as
\begin{equation}
    \frac{d \Tilde{\rho}(t)}{dt} = -i[\Tilde{H}_{I}(t), \Tilde{\rho}(t)],
\end{equation}
where a tilde denotes an operator expressed in the interaction picture. We can then introduce the Liouvillian operator such that
\begin{equation}
    \frac{d\Tilde{\rho}(t)}{dt} = \mathcal{L}(t)\Tilde{\rho}(t),
    \label{Time_Evolution_Interaction}
\end{equation}
where  the Liouvillian is defined as $\mathcal{L}=-i[\Tilde{H}_{I}(t), \bullet ]$. Eq.~\eqref{Time_Evolution_Interaction} has the general solution,
\begin{equation}
    \Tilde{\rho}(t) = \overleftarrow{{T}}\exp\big(\int_{0}^{t}ds\, \mathcal{L}(s)\big)\Tilde{\rho}(0),
\end{equation}
where $\overleftarrow{T}$ is the time-ordering operator. If we are only interested in the dynamics of the system, then we can take the trace of the bath, assuming initial conditions $\Tilde{\rho}(0) = \Tilde{\rho}_{S}(0) \otimes \Tilde{\rho}_{\beta} $ where $\Tilde{\rho}_{\beta}$ is an arbitrary initial state of the anharmonic mode, and $\Tilde{\rho}_{S}(0)$ is the initial state of the system. 
Taking the partial trace over the environment yields
\begin{equation}
    \Tilde{\rho}_{S}(t) = \bigg(\bigg\langle \overleftarrow{T}\exp\big(\int_{0}^{t}ds\, \mathcal{L}(s)\big)\bigg\rangle_\beta \bigg)\Tilde{\rho}_{S}(0).
\end{equation}
We next introduce that the environment can be made up of $N$ discrete anharmonic modes such that $\Tilde{H}_{I}(t) = \sum_{\alpha = 0}^{N-1}S(t) \otimes g(\alpha)B_{\alpha}(t)$ such that the reduced density matrix takes the form 
\begin{equation}
    \Tilde{\rho}_{S}(t) = \bigg(\bigg\langle\overleftarrow{T}\exp\big(\int_{0}^{t}ds\sum_{\alpha=0}^{N-1} \mathcal{L}_{\alpha}(s) \big)\bigg\rangle_\beta\bigg)\Tilde{\rho}_{S}(0)
    \label{Influence_Functional_Reduced_System},
\end{equation}
where $\mathcal{L}_{\alpha}(s) = -i[S(s) \otimes g(\alpha)B_{\alpha}(s),\bullet]$. To show that the influence of a non-Gaussian environment reduces to a Gaussian form in the thermodynamic limit, we consider a cumulant expansion of the influence functional. The influence function,
\begin{equation}
    F(t) = \bigg(\langle\overleftarrow{T}\exp\big(\int_{0}^{t}ds \sum_{\alpha = 0}^{N-1}\mathcal{L}_{\alpha}(s)\big) \rangle_\beta\bigg),
\end{equation}can be expanded as a Dyson series in powers of the interaction Liouvillian, resulting in Eq.~\eqref{Influence_Functional_Reduced_System} being expressed as,
\begin{equation}
\Tilde{\rho}_{S}(t)=\sum_{n=0}^{\infty}\frac{1}{n!}
 \left\langle\overleftarrow{T}\left[\int_{0}^{t} ds 
\sum_{\alpha = 0}^{N-1}\mathcal{L}_{\alpha}(s)
 \right]^n\right\rangle_\beta \Tilde{\rho}_{S}(0).
\label{Dysonexpansion}
\end{equation}
In order to express Eq.~\eqref{Dysonexpansion} as a cumulant expansion over the anharmonic degrees of freedom, we redefine the Liouvillian, $\mathcal{L}_{\alpha}$, as
\begin{equation}
\begin{aligned}
    \mathcal{L}_{\alpha}(s) = & -i[S(s) \otimes g(\alpha)B_{\alpha}(s), \bullet] \\&=
    -i\sum_{d = \pm} \big\{\hat{\hat{S}}^{d}(s) \otimes\hat{\hat{B}}^{d}_{\alpha}(s) \big\} \bullet,  
\end{aligned}
\end{equation}
where the super-operators are defined as
\begin{equation}
\begin{aligned}
    \hat{\hat{S}}^{+}(s) \bullet = S(s) \bullet, \hat{\hat{S}}^{-}(s) \bullet= -\bullet S(s)&, &\\ \hat{\hat{B}}_{\alpha}^{+}(s) \bullet = g(\alpha)B(s) \bullet, \hat{\hat{B}}_{\alpha}^{-}(s) \bullet = \bullet g(\alpha) B(s).
\label{Super_Operator}
\end{aligned}
\end{equation}
Given that the time-ordering operator can be decomposed into system and environment time-ordering operators ($\overleftarrow{T} (S(t)\otimes g(\alpha)B_{\alpha}(t)) = \overleftarrow{T}_S(S(t))\otimes\overleftarrow{T}_{B}(g(\alpha)B_{\alpha}(t))$) then Eq.~\eqref{Dysonexpansion} can be written as 
\begin{equation}
\begin{aligned}
\widetilde{\rho}_S(t)=\bigg({}&\sum_{n=0}^{\infty}\frac{(-i)^n}{n!}
\int_0^t ds_1\cdots\int_0^t ds_n \\& \times 
\sum_{d_1,\ldots,d_n} \sum_{\alpha_{1}, \ldots \alpha_{n}}^{N-1}
C^{d_1,\ldots,d_n}_{\alpha_{1}, \ldots, \alpha_{n}}(s_1,\ldots,s_n)
\\
&\qquad\times
\overleftarrow{T}_S
\left[
\hat{\hat S}^{d_1}(s_1)\cdots
\hat{\hat S}^{d_n}(s_n)
\right]\bigg)\Tilde{\rho}_{S}(0).
\label{MomentExpansion}
\end{aligned}
\end{equation}
The n-th order bath correlation function is defined as
\begin{equation}
    C^{d_{1}, \ldots , d_{n}}_{\alpha_{1}, \ldots \alpha_{n}}(s_{1}, \ldots s_{n}) = \langle \overleftarrow{T}_{B} \hat{\hat{B}}_{\alpha_{1}}^{d_{1}}(s_{1}) \ldots \hat{\hat{B}}_{\alpha_{n}}^{d_{n}}(s_{n})\rangle .
    \label{Moments}
\end{equation}
If we impose that the environments are statistically independent such that their joint state factorizes as $\rho_{\beta} = \bigotimes_{\alpha} \rho_{\beta_{\alpha}}$, then it follows that
\begin{equation}
\begin{aligned}
    &\sum_{\substack{
\alpha_1,\ldots,\alpha_n=0\\
\alpha_i\neq\alpha_j\;\forall\,i\neq j
}}^{N-1}\langle \hat{\hat{B}}^{d_{1}}_{\alpha_{1}}(s_{1}) \ldots \hat{\hat{B}}^{d_{n}}_{\alpha_{n}}(s_n)\rangle_{\beta} = \sum_{\alpha_{i} = 0}^{N-1}\bigg[\prod_{i=1}^{N} \langle \hat{\hat{B}}_{\alpha_{i}}^{d_{i}}(s_{i})\rangle_{\beta_{\alpha_{i}}}\bigg].
\end{aligned}
\end{equation}
This has an important implication once we reformulate Eq.~\eqref{MomentExpansion} as a cumulant expansion over the anharmonic degrees of freedom using the generalised cumulant theorem ~\cite{kubo_generalized_1962}. The cumulant generating function is defined as
\begin{equation}
    \exp(K(t)) =  F(t),
\end{equation}
where $K(t)$ is the cumulant generating function and $F(t)$ is the influence functional. The cumulants $k_{n}$ can then be found through a power series expansion of the cumulant generating function
\begin{equation}
    K(t) = \sum_{i=1}^{\infty}\frac{(-i)^{n}}{n!}k_{n}(t),
\end{equation}
where $k_{n}(t)$ is given by 
\begin{equation}
\begin{aligned}
k_{n}(t)=&\bigg({}
\int_0^t ds_1\cdots\int_0^t ds_n \\& \times 
\sum_{d_1,\ldots,d_n} \sum_{\alpha_{1}, \ldots \alpha_{n}}^{N-1}
[C^{d_1,\ldots,d_n}_{\alpha_{1}, \ldots, \alpha_{n}}(s_1,\ldots,s_n)]_{C}
\\
&\qquad\times
\overleftarrow{T}_S
\left[
\hat{\hat S}^{d_1}(s_1)\cdots
\hat{\hat S}^{d_n}(s_n)
\right]\bigg).
\end{aligned}
\end{equation}
The subscript $C$ denotes the connected (cumulant) part of the correlation function such that 
\begin{equation}
    [C^{d_1,\ldots,d_n}_{\alpha_{1}, \ldots, \alpha_{n}}(s_1,\ldots,s_n)]_{C} = \langle \overleftarrow{T}_{B} \hat{\hat{B}}_{\alpha_{1}}^{d_{1}}(s_{1}) \ldots \hat{\hat{B}}_{\alpha_{n}}^{d_{n}}(s_{n})\rangle_{C}.
\end{equation}
Given that we can reformulate the influence functional as a cumulant expansion Eq.~\eqref{MomentExpansion} can be re-expressed as,
\begin{equation}
\begin{aligned}
    \widetilde{\rho}_{S}(t)
    = \overleftarrow{T}
    & \exp\bigg \{
    \sum_{n=1}^{\infty}
    \frac{(-i)^{n}}{n!}
    \int_{0}^{t} ds_{1} \ldots \int_{0}^{t}ds_{n} \\
    &\times
    \sum_{d_{1},\ldots d_{n}}\sum_{\alpha_{1}, \ldots, \alpha_{n} =0}^{N-1}
    \left\langle
    \overleftarrow{T}_{B}
   \hat{\hat{B}}_{\alpha_{1}}^{d_{1}}(s_{1})
    \ldots
    \hat{\hat{B}}_{\alpha_{n}}^{d_{n}}(s_{n})
    \right\rangle_{\beta,C} \\
    &\times
    \overleftarrow{T}_{S}
    \left[
    \hat{\hat{S}}^{d_{1}}(s_{1})
    \ldots
    \hat{\hat{S}}^{d_{n}}(s_{n})
    \right]
    \bigg \}
    \widetilde{\rho}_{S}(0).
\end{aligned}
\end{equation}
Since the anharmonic environments are statistically independent, all mixed cumulants involving operators from different baths vanish. Thus, the only surviving terms of the cumulant expansion can be expressed as, 
\begin{equation}
\begin{aligned}
&\sum_{d_1,\ldots,d_n}\sum_{\alpha}^{N-1}\langle
    \overleftarrow{T}_{B}
    \hat{\hat{B}}_{\alpha}^{d_{1}}(s_{1})
    \ldots
    \hat{\hat{B}}_{\alpha}^{d_{n}}(s_{n})
    \rangle_{\beta,C} \\& = \sum_{d_1,\ldots,d_n}\sum_{\alpha_1, \dots, \alpha_N}^{N-1}  \prod_{i=1}^{N-1} \delta_{\alpha_i, \alpha_{i+1}} \langle T_{B}\hat{\hat{B}}_{\alpha_{1}}^{d_{1}}(s_{1}) \ldots \hat{\hat{B}}_{\alpha_{n}}^{d_{n}}(s_{n}) \rangle_{\beta,C}.
\end{aligned}
\end{equation}
Consequently, the $n$-th order cumulant expansion is reduced to
\begin{equation}
\begin{aligned}
    \widetilde{\rho}_{S}(t)
    = \overleftarrow{T}
    \exp\bigg \{
    &\sum_{n=1}^{\infty}
    \frac{(-i)^{n}}{n!}
    \int_{0}^{t} ds_{1} \ldots ds_{n} \\
    &\times
    \sum_{d_{1}, \ldots, d_{n}}\sum_{\alpha =0}^{N-1}
    \left\langle
    \overleftarrow{T}_{B}
    \hat{\hat{B}}_{\alpha}^{d_{1}}(s_{1})
    \ldots
    \hat{\hat{B}}_{\alpha}^{d_{n}}(s_{n})
    \right\rangle_{\beta,C} \\
    &\times
    \overleftarrow{T}_{S}
    \left[
    \hat{\hat{S}}^{d_{1}}(s_{1})
    \ldots
    \hat{\hat{S}}^{d_{n}}(s_{n})
    \right]
    \bigg \} 
    \widetilde{\rho}_{S}(0).
\end{aligned}
\end{equation}
Now assume that the system-bath coupling scales as $g(\alpha) \propto 1/\sqrt{N}$, which can occur when discretising a bath through its spectral density and which naturally arises in photonic ~\cite{milonni_introduction_2019} and phononic~\cite{HOLSTEIN1959325} bath couplings, where local modes are decomposed into momentum space. Also, assuming that the bath operator has zero mean, which can be done through a simple displacement transformation such as Eq.~\eqref{Eq:appropriate_coupling}. Therefore, the cumulant expansion of the $n$-th order becomes,
\begin{equation}
\begin{aligned}
    \sum_{d_{1}, \ldots, d_{n}}\sum_{\alpha =0}^{N-1}
    \left\langle
     \overleftarrow{T}_{B}
    \hat{\hat{B}}_{\alpha}^{d_{1}}(s_{1})
    \ldots
    \hat{\hat{B}}_{\alpha}^{d_{n}}(s_{n})
    \right\rangle_{\beta,C}  \propto{N^{1-n/2}} 
\end{aligned}
\end{equation}
as each $\hat{\hat{B}}$ contains the system-bath coupling strength $\propto N^{-1/2}$, see equation Eq.~\eqref{Super_Operator}. Since the sum  of the $n$-th cumulant expansion contains $\mathcal{O}$($N$) independent contributions, the $n$-th order cumulants scale as $\mathcal{O}(N^{1-n/2})$. Consequently, in the thermodynamic limit as the number of anharmonic modes goes to infinity ($N \rightarrow \infty$), all cumulant terms higher than order two go to zero, i.e. $1/N^{(n/2)}\times \sum_{d_{1}, \ldots, d_{n}}\sum_{\alpha =0}^{N-1} \langle(\overleftarrow{T}_{B}
    \hat{\hat{B}}_{\alpha}^{d_{1}}(s_{1})
    \ldots
    \hat{\hat{B}}_{\alpha}^{d_{n}}(s_{n})\rangle_{C} \rightarrow 0$ for $n > 2)$. This then leaves the second-order cumulant expansion
\begin{equation}
\begin{aligned}
    \widetilde{\rho}_{S}(t)
    = \overleftarrow{T}
    \exp\bigg \{
    &\frac{(-i)^{2}}{2}
    \int_{0}^{t} ds_{1}\,\int_{0}^{t}ds_{2}
    \\&\times\sum_{d_{1}, d_{2}}\sum_{\alpha =0}^{N-1}
    \left\langle
    \overleftarrow{T}
    \hat{\hat{B}}_{\alpha}^{d_{1}}(s_{1})
    \hat{\hat{B}}_{\alpha}^{d_{2}}(s_{2})
    \right\rangle_{\beta,C}
    \\& \times \overleftarrow{T}_{S}
    \left[
    \hat{\hat{S}}^{d_{1}}(s_{1})
    \hat{\hat{S}}^{d_{2}}(s_{2})
    \right]
    \bigg \}
    \widetilde{\rho}_{S}(0).
    \label{Cumulant_Expansion_Second_Order}
\end{aligned}
\end{equation}
Using the generalized cumulant theorem, the second-order cumulant expansion can be expressed in terms of the first- and second-order moments. Consequently, the dynamics of the system are completely determined by the two-time correlation functions of the anharmonic bath, leading to Eq.~\eqref{Cumulant_Expansion_Second_Order} being rewritten as,
\begin{equation}
\begin{aligned}
    \widetilde{\rho}_{S}(t)
    = \overleftarrow{T}
    \exp\bigg[
    &\frac{(-i)^{2}}{2}
    \int_{0}^{t} ds_{1}\,\int_{0}^{t}ds_{2}
    \sum_{d_{1}, d_{2}}\sum_{\alpha =0}^{N-1}C^{d_{1},d_{2}}_{\alpha}(s_{1},s_{2})
    \\
    &\times
    \overleftarrow{T}_{S}
    \left[
    \hat{\hat{S}}^{d_{1}}(s_{1})
    \hat{\hat{S}}^{d_{2}}(s_{2})
    \right]
    \bigg]
    \widetilde{\rho}_{S}(0).
    \label{Final_Cumulant}
\end{aligned}
\end{equation}
Thus, with the chosen system-bath coupling scaling and within the thermodynamic limit, the influence functional for an anharmonic bath reduces to an expression that depends solely on the second-order cumulant, and hence solely on the two-point bath correlation function. Within this limit, all higher-order cumulants vanish, such that the influence functional takes a Gaussian form. Following from this, we can express Eq.~\eqref{Final_Cumulant} using the standard Feynman-Vernon path integral formalism~\cite{feynman_theory_1963} through first expanding out
\begin{equation}
\begin{aligned}
    \sum_{d_{1},d_{2}}\sum_{\alpha = 0}^{N-1}C_{\alpha}^{d_{1},d_{2}}(s_{1},s_{2})
    \times \overleftarrow{T}_{S}[\hat{\hat{S}}^{d_{1}}(s_{1})\hat{\hat{S}}^{d_{2}}(s_{2})] &\\
    = \sum_{\alpha =0}^{N-1}\bigg\{C_{\alpha}^{+,+}(s_{1},s_{2}) \times \overleftarrow{T}_{S}[\hat{\hat{S}}^{+}(s_{1})\hat{\hat{S}}^{+}(s_{2})] &\\
    + C_{\alpha}^{+,-}(s_{1},s_{2}) \times \overleftarrow{T}_{S}[\hat{\hat{S}}^{+}(s_{1})\hat{\hat{S}}^{-}(s_{2})] &\\
    + C_{\alpha}^{-,+}(s_{1},s_{2}) \times \overleftarrow{T}_{S}[\hat{\hat{S}}^{-}(s_{1})\hat{\hat{S}}^{+}(s_{2})] &\\
    + C_{\alpha}^{-,-}(s_{1},s_{2}) \times \overleftarrow{T}_{S}[\hat{\hat{S}}^{-}(s_{1})\hat{\hat{S}}^{-}(s_{2})]\bigg\}.
    \label{ExpandedOutFeynmannVernon}
\end{aligned}
\end{equation}
Therefore, Eq.~\eqref{ExpandedOutFeynmannVernon} can be expressed as 
\begin{equation}
\begin{aligned}
    \widetilde{\rho}_{S}(t)
    = \overleftarrow{T}
    \exp\bigg[
    &-
    \int_{0}^{t} ds_{1}\,\int_{0}^{s_{1}}ds_{2}\, O^{C}(s)\bigg[C_{R}(s - s') O^{C}(s') \\& + iC_{I}(s - s')O^{a}(s')]\bigg]\widetilde{\rho}_{S}(0),
\end{aligned}
\end{equation}
where $C_{R}(s - s')$ and $C_{I}(s -s')$ denote the real and imaginary parts of the two-time bath correlation function respectively and $O^{c}(s) = [O(s), \bullet]$ and $O^{a}(s) = \{O(s), \bullet\}$ are the commutator and anti-commutator respectively.

\section{Derivation of Effective Spectral Density for a continuous anharmonic environment}\label{App:ContSpecDens}

The correlation function takes the form
\begin{align}
    \mathcal{C}(t)  = & \int_0^\infty d\alpha J(\alpha)\big(\sum_{n\neq m =0}^{N_{\alpha}-1} |B_{nm}(\alpha)|^2e^{iE_{nm}(\alpha)t}p_n(\alpha)+\Delta D_\alpha^2\big).
\end{align}
The associated effective spectral density is 
\begin{align}
     J_\text{eff}(\omega,\beta) &=\frac{1}{\pi}\tanh (\beta\omega/2) \int^\infty_0 \text{Re}\{\mathcal{C}(t)\}\cos (\omega t) dt\\&=\frac{1}{2\pi}\tanh (\beta\omega/2) \int^\infty_0 \text{Re}\{\mathcal{C}(t)\}(e^{i\omega t}+e^{-i\omega t})dt \nonumber.
\end{align}

Defining a new function 
\begin{equation}
    X_{nm}(\alpha)= J(\alpha) |B_{nm}(\alpha)|^2p_n(\alpha)
\end{equation}
such that 
\begin{equation}
    \mathcal{C}(t) = \int_0^\infty d\alpha (\sum_{n\neq m=0}^{N-1}X_{nm}(\alpha)e^{iE_{nm}(\alpha)t} +J(\alpha)\Delta D_\alpha^2).
\end{equation}
% and we assume that all oscillators have equivalent energy levels up to a rescaling by $\alpha$ such that $E_{nm}(\alpha)=E_{nm}\alpha$, then 
% \begin{equation}
%     e^{iE_{nm}(\alpha)t} = e^{iE_{nm}\alpha t}
% \end{equation}

From this, let us consider the rate functions of this 
\begin{align}
    \gamma(\nu) =\Re{\int_0^\infty dt e^{-i\nu t}\mathcal{C}(t)}
\end{align}
we can utilise the Sokhotski–Plemelj theorem to get 
\begin{equation}
    \gamma(\nu)= \pi \int_0^\infty d\alpha \sum_{n\neq m=0}^{N_{\alpha}-1}X_{nm}(\alpha)\delta(\nu-E_{nm}(\alpha))+\pi\delta(\nu)A_\text{v}
\end{equation}
with $A_\text{v}=\int_0^\infty d\alpha J(\alpha)\Delta D^2_\alpha$.
Focusing on the first term, we find that 
\begin{align}
   & \int_0^\infty d\alpha \sum_{n\neq m=0}^{N_{\alpha}-1}X_{nm}(\alpha)\delta(\nu-E_{nm}(\alpha))
    \\&=\sum_{n\neq m=0}^{N-1}\sum_{i}\frac{X_{mn}(\alpha_{i})}{|{E'_{nm}(\alpha_{i})|}},\nonumber
\end{align}
where the $\alpha_i$ are the roots of the function $f(\alpha) = \nu - E_{nm}(\alpha)$ and $E'_{nm}(\alpha) = \frac{dE_{nm}(\alpha_{i})}{d\alpha}$. We can also note that 
\begin{equation}
    X_{nm}(\alpha)=0 \text{  for } \alpha<0,
\end{equation}
as the spectral density has no support on the negative real line.
As such as $E_{nm}(\alpha)<0$ when $n<m$
\begin{equation}
    \gamma(\nu) = \begin{cases}
        \pi \sum_{n > m = 0}^{N_{\alpha} -1}\sum_{i} \frac{X_{mn}(\alpha_{i})}{|{E'_{nm}(\alpha_{i})|}} \text{ for }\nu > 0 \\
        \pi \sum_{n < m = 0}^{N_{\alpha} -1}\sum_{i} \frac{X_{mn}(\alpha_{i})}{|{E'_{nm}(\alpha_{i})}|} \text{ for }\nu < 0
    \end{cases}
\end{equation}
by noting that $p_n(\alpha_{i})=e^{-\beta(E_{nm}(\alpha_{i}))}p_m(\alpha_{i})$ for $n>m$ and that $B_\alpha$ is Hermitian, we have the detailed balance equations
\begin{equation}
    \gamma(\omega)=e^{-\beta\omega}\gamma(-\omega).
\end{equation}

It follows from the definition of the effective spectral density that for non-negative $\omega$ that
\begin{align}
    J_\text{eff}(\omega,\beta) &= \frac{1}{2\pi}\tanh(\beta\omega/2)(\gamma(\omega)+\gamma(-\omega))\\&= \frac{1}{2\pi}\tanh(\beta\omega/2)(1+e^{-\beta \omega})\gamma(\omega)
   \nonumber \\&=\frac{1}{2}(1-e^{-\beta\omega})\times   \nonumber \\&\sum_{n\neq m=0}^{N-1}\sum_{i} \frac{X_{mn}(\alpha_{i})}{|{E'_{nm}(\alpha_{i})|}}  + A_\text{v}\delta(\omega).
  \nonumber \\ &=\frac{1}{2}(1-e^{-\beta\omega})\times\nonumber\\ &\sum_{n>m=0}^{N-1}\sum_{i}\frac{|B_{nm}(\alpha_{i})|^2}{|{E'_{nm}(\alpha_{i})|}}J(\alpha_{i})p_n(\alpha_{i}) \nonumber\\&~+A_\text{v}\delta(\omega),\nonumber
\end{align}
if we now assume that the AMs have the same energy level up to a linear scaling by $\alpha$ (i.e. $E_{nm}(\alpha) = \alpha E_{nm}$) then we return to the same form as in the main text,
\begin{align}
    J_\text{eff}&(\omega,\beta) =\frac{1}{\pi}\tanh (\beta\omega/2) \int^\infty_0 \text{Re}\{\mathcal{C}(t)\}\cos (\omega t) dt\,\nonumber\\ =&\frac{1}{2}(1-e^{-\beta\omega})\times\nonumber\\ &\sum_{n>m = 0}^{N_{\alpha} -1} |B_{nm}\bigg(\frac{\omega}{E_{nm}}\bigg)|^2J\bigg(\frac{\omega}{E_{nm}}\bigg)
    \frac{1}{E_{nm}}p_n\bigg(\frac{\omega}{E_{nm}}\bigg) \nonumber\\&\hspace{20pt}+A_\text{v}\delta(\omega).
\end{align}
\section{Derivation of Lindblad master equation for Damped oscillator}\label{App:DampME}
Consider a single AM coupled to a harmonic oscillator bath via the coupling operator $B$, we can decompose the Hamiltonian for such a system into a free and an interaction component in the following way
\begin{align}
H_F &= \sum_{n = 0}^{N_{\alpha} -1} E_n \ket{n}\bra{n} + \sum_\lambda \omega_\lambda a^\dag_\lambda a_\lambda, \\
H_I &=  \sum_{nm = 0}^{N_{\alpha} -1}B_{nm}\ket{n}\bra{m}(\sum_\lambda g_\lambda(a_\lambda+a^\dag_\lambda))
\end{align}
in the interaction picture, this then yields the time-dependent interaction Hamiltonian
\begin{equation}
    \Tilde{H}_I =  \sum_{nm=0}^{N_{\alpha} -1}B_{nm}\ket{n}\bra{m}e^{-i\omega_{nm}t}(\sum_\lambda g_\lambda(a_\lambda e^{i\omega_\lambda t}+a^\dag_\lambda e^{-i\omega_\lambda t})).
\end{equation}
Via the time-convolutionless second-order master equation, we have after making the second Markov approximation
\begin{equation}
    \frac{d}{dt}\Tilde{\rho}_S = -\Tr_E\big\{\int_0^\infty ds [\Tilde{H}_I(t),[\Tilde{H}_I(t-s),\Tilde{\rho}_S(t)\otimes\rho_\beta]]\big\}
\end{equation}
inside the trace we have four unique operators 
\begin{align}
    \Tilde{H}_{I}(t)\Tilde{H}_{I}(t-s)\Tilde{\rho}_S(t)\otimes\rho_\beta, \\ \nonumber
    -\Tilde{H}_{I}(t)\Tilde{\rho}_S(t)\otimes\rho_\beta \Tilde{H}_{I}(t-s),\\ \nonumber
    -\Tilde{H}_{I}(t-s)\Tilde{\rho}_S(t)\otimes\rho_\beta \Tilde{H}_{I}(t),\\ \nonumber
    +\Tilde{\rho}_S(t)\otimes\rho_\beta \Tilde{H}_{I}(t-s)\Tilde{H}_{I}(t). 
\end{align}
Performing the partial trace where $C(t)$ is the bath correlation function and expanding the operators into their components yields
\begin{align}
    & C(s)(\sum_{nm}B_{nm}\ket{n}\bra{m}e^{-i\omega_{nm}t})\\& \nonumber \times (\sum_{kl}B_{kl}\ket{k}\bra{l}e^{-i\omega_{kl}(t-s)})\Tilde{\rho}_S(t),\\& \nonumber
    -C^*(s)(\sum_{nm}B_{nm}\ket{n}\bra{m}e^{-i\omega_{nm}t}) \\& \nonumber \times\Tilde{\rho}_S(t)(\sum_{kl}B_{kl}\ket{k}\bra{l}e^{-i\omega_{kl}(t-s)}),\\& \nonumber
    -C(s)(\sum_{kl}B_{kl}\ket{k}\bra{l}e^{-i\omega_{kl}(t-s)}) \\&  \nonumber\times\Tilde{\rho}_S(t)(\sum_{nm}B_{nm}\ket{n}\bra{m}e^{-i\omega_{nm}t}),\\&  \nonumber
    C^*(s)\Tilde{\rho}_S(t)(\sum_{kl}B_{kl}\ket{k}\bra{l}e^{-i\omega_{kl}(t-s)}) \\& \nonumber \times(\sum_{nm}B_{nm}\ket{n}\bra{m}e^{-i\omega_{nm}t}).
\end{align}
Resolving the outer integral yields 
\begin{align}
    &\Gamma(\omega_{kl})(\sum_{nm}B_{nm}\ket{n}\bra{m}e^{-i\omega_{nm}t}) \\& \nonumber \times(\sum_{kl}B_{kl}\ket{k}\bra{l}e^{-i\omega_{kl}t})\Tilde{\rho}_S(t) \\& \nonumber
    -\Gamma^*(\omega_{kl})(\sum_{nm}B_{nm}\ket{n}\bra{m}e^{-i\omega_{nm}t}) \\& \nonumber \times \Tilde{\rho}_S(t)(\sum_{kl}B_{kl}\ket{k}\bra{l}e^{-i\omega_{kl}t}) \\& \nonumber 
    -\Gamma(\omega_{kl})(\sum_{kl}B_{kl}\ket{k}\bra{l}e^{-i\omega_{kl}t}) \\& \nonumber \times \Tilde{\rho}_S(t)(\sum_{nm}B_{nm}\ket{n}\bra{m}e^{-i\omega_{nm}t})\\& \nonumber 
    \Gamma^*(\omega_{kl})\Tilde{\rho}_S(t)(\sum_{kl}B_{kl}\ket{k}\bra{l}e^{-i\omega_{kl}t}) \\& \nonumber \times (\sum_{nm}B_{nm}\ket{n}\bra{m}e^{-i\omega_{nm}t}).
\end{align}
We can see that, in the interaction picture, for some terms time-dependent phases remain; as such, we perform a secularisation such that only like terms remain, that is $m=k $ and $n=l$ such that $\omega_{nm}=-\omega_{kl}$. This then leaves
\begin{align}
    \sum_{nm}\Gamma(\omega_{mn})B_{nm}^2\ket{n}\bra{m}\ket{m}\bra{n}\Tilde{\rho}_S(t)\\ \nonumber
    -\sum_{nm}\Gamma^*(\omega_{mn})B^2_{nm}\ket{n}\bra{m}\Tilde{\rho}_S(t)\ket{m}\bra{n}\\ \nonumber
    -\sum_{nm}\Gamma(\omega_{mn})B_{mn}^2\ket{m}\bra{n}\Tilde{\rho}_S(t)\ket{n}\bra{m}\\ \nonumber 
    \sum_{nm}\Gamma^*(\omega_{mn})B^2_{mn}\Tilde{\rho}_S(t)\ket{m}\bra{n}\ket{n}\bra{m}. \nonumber 
\end{align}
Neglecting Lamb shifts, this has the Lindblad form 
\begin{equation}
    \frac{d}{dt}\Tilde{\rho}_{s} = \sum_{nm = 0}^{N_{\alpha} -1}\mathcal{L}(\Gamma(\omega_{mn})B_{nm}^2,\ket{n}\bra{m}).
\end{equation}
If we assume that the spectral density of the harmonic oscillator environment is locally flat across the transition energies of the AM and zero at the origin 
\begin{equation}
    \Gamma(\omega) = \begin{cases}
        \gamma(n(\omega)+1), \hspace{5pt} \omega>0,\\
        \gamma n(\omega), \hspace{5pt} \omega<0,\\
        0, \hspace{5pt} \omega=0,\\
    \end{cases}
\end{equation}
yielding the Lindblad master equation described in the main text.

\section{Derivation of effective spectral density for a single damped anharmonic oscillator}\label{App:DampSpecDens}
Through application of the Lindblad master equation, we can derive a set of differential equations for the elements $R_{nm}$,
\begin{align}
\dot{R}_{nm}(t) &= \left(
    -i E_{\alpha, nm}
    - \frac{1}{2} \big(
        \gamma_n^{\uparrow}
        + \gamma_n^{\downarrow}
        + \gamma_m^{\uparrow}
        + \gamma_m^{\downarrow}
    \big)
\right) R_{nm}(t) \nonumber \\
&\quad
+ \sum_{s,p} \gamma^{\downarrow}_{ns}
    L^{\downarrow}_{nm,sp} \, R_{sp}(t)
+ \sum_{q,p} \gamma^{\uparrow}_{mq}
    L^{\uparrow}_{nm,qp} \, R_{qp}(t)
    \label{Eq:dotCNM(t)},
\end{align}
where
\begin{align}
\gamma_n^{\uparrow} &= \sum_{s < n} |B_{ns}|^{2} \gamma_{ns}^{\uparrow}, \quad
\gamma_n^{\downarrow} = \sum_{s > n} |B_{ns}|^{2} \gamma_{ns}^{\downarrow}, \\
\gamma_m^{\uparrow} &= \sum_{s < m} |B_{ms}|^{2} \gamma_{ms}^{\uparrow}, \quad
\gamma_m^{\downarrow} = \sum_{s > m} |B_{ms}|^{2} \gamma_{ms}^{\downarrow}.
\end{align}
These quantities represent the total upward (absorption) and downward (decay) transition rates associated with the resonant energy levels $n$ and $m$, whereas the coefficients $L_{nm,sp}^{\downarrow} = \sum_{s>m}\sum_{p>n}B_{ns}B_{pm}$ and $L_{nm,qp}^{\uparrow} = \sum_{q<n}\sum_{p<m} B_{nq}B_{pm}$ account for decay and absorption processes for pairs of off-resonant energy levels that are not $n$ and $m$. Here we split the bath correlation function into diagonal ($n = m$) and off-diagonal components ($n \neq m$). First, we consider the off-diagonal components, for which we make an initial approximation to remove the dependence of the off-resonant decay and absorption process corresponding to $R_{sp}(t)$ and $R_{qp}(t)$. This effectively takes a secular approximation, since these transitions will have negligible effects compared to the resonant transitions from $R_{nm}(t)$, leading to
\begin{align}
\dot{R}_{nm}(t) &\approx \left(
    -i E_{\alpha, nm}
    - \frac{1}{2} \gamma_{nm}
\right) R_{nm}(t),
\label{dotC}
\end{align}
where all the transition rates have been combined into one function $\gamma_{nm} = \gamma_{n}^{\uparrow} +\gamma_{m}^{\uparrow} + \gamma_{n}^{\downarrow} + \gamma_{m}^{\downarrow}$ and $E_{\alpha, nm} = E_n(\alpha) - E_m(\alpha)$.
Solving Eq.~\eqref{dotC} leads to the trivial solution,  
\begin{equation}
    {R}_{nm}(t) = R_{nm}(0)\exp\left\{- i E_{\alpha, nm}t
   - \frac{1}{2}\gamma_{nm}t\right\}.
\end{equation}

We wish to calculate the Fourier transform of this bath correlation function, given by 
\begin{equation}
\begin{split}
\Gamma(\nu) &= \int dt\, e^{-i\nu t} C(t) \\
&= \sum_{nm=0}^{N-1} \int dt\, B_{\alpha,nm}
R_{nm}(t) e^{-i\nu t}.
\end{split}
\end{equation}
Through the use of the Laplace transformation, we can convert the differential equation Eq.~\eqref{dotC} to the form,
\begin{equation}
    \Tilde{R}_{nm}(s) \approx \frac{R_{nm}(0)}{s + i\omega_{nm} + \gamma_{nm}}.
\end{equation}
This then enables the calculation of the real part of the Fourier transform to be given as 
\begin{equation}
    \Re({\Gamma_{nm}(\omega)}) \approx \frac{|x_{nm}|^{2} p_{n}\gamma_{nm}}{(\omega + E_{\alpha, nm})^{2} + \gamma_{nm}^{2}},
    \label{nnotm}
\end{equation}
where the $\Gamma_{nm}(\omega)$ are the matrix elements of $\Gamma(\omega)$ such that $\Gamma(\nu) = \sum_{n,m=0}^{N =0}\Gamma_{n,m}(\nu)$. The framework reveals a similarity to the continuous case where the spectral functions of the damped anharmonic oscillator correspond to Lorentzian peaks centred around transition frequencies within the anharmonic oscillator. The line widths of these peaks are determined by the sum of the corresponding decay and absorption rates for each transition. Notably, these rates exhibit a strong temperature dependence, leading to a broadening of the peaks at larger temperatures that was not present in the continuous case.\\

In addition, we compute the diagonal elements of the bath correlation function, explicitly accounting for off-resonant transitions, which can contribute significantly alongside decay and absorption processes. The bath correlation function can then be rewritten as,
\begin{equation}
    \dot{R}_{nn}(t) = -(\gamma^{\uparrow}_{n} + \gamma^{\downarrow}_{n})R_{nn} + \sum_{s \neq n}\gamma_{ns}R_{ss} ,
    \label{eq:n=n}
\end{equation}
where $\gamma_{ns} = \sum_{s < n} |B_{ns}|^{2} \gamma_{ns}^{\uparrow} + \sum_{s > n} |B_{ns}|^{2} \gamma_{ns}^{\downarrow}$ and approximated the off-resonant transitions within Eq.~\eqref{Eq:dotCNM(t)} with a secular approximation such that $R_{sp} = R_{ss}$. This linear differential equation can then be solved, given that 
\begin{equation}
    \dot{R}_{nn} = \sum_{s}M_{ns} R_{ss},
\end{equation}
where $M_{ns}$ is the rate matrix that accounts for the population dynamics, given by
\begin{equation}
M_{ns} =
\begin{cases}
\gamma_{ns}, & n \neq s, \\
-\displaystyle(\gamma_{n}^{\uparrow} + \gamma_{n}^{\downarrow}), & n = s.
\end{cases}
\end{equation}
Introducing vectorised notation, we can reformulate the equation as
\begin{equation}
    \dot{R}(t) = M R
\end{equation}
diagonalising the rate matrix $M$, we write $V^{-1}M V = D$, where $D_{ii} = \lambda_{i}$ are the eigenvalues and the columns of $V$ are the eigenvectors. Transforming into the diagonal basis, $K(t) = V^{-1}R(t)$, yields the formal solution,
\begin{equation}
    K_{i}(t) = e^{-\lambda_{i} t} K_{i}(0).
\end{equation}
We can then use this to find the spectral function given in the form,
\begin{equation}
    \Re(\Gamma_{D}(\nu)) =\sum_n \Re(\Gamma_{nn}(\nu)) \approx \sum_{i = 0} \frac{{X}_{i}\lambda_{i}}{\nu^{2} + \lambda_{i}^{2}}
    \label{n=m},
\end{equation}
where ${X}_i$ are the weights given by
$
{X_{i}} =\sum_{n}V_{ni} K_{i}(0)$.
Here, we find that the damped anharmonic mode has significant Lorentzian structure centred around zero frequency, with the peak widths determined by the eigenvalue $\lambda_{i}$.\\

\bibliography{main}

@article{caldiera1983path,
title = {Path integral approach to quantum Brownian motion},
journal = {Phys. A: Stat. Mech. Appl.},
volume = {121},
number = {3},
pages = {587-616},
year = {1983},
issn = {0378-4371},
doi = {10.1016/0378-4371(83)90013-4},
author = {A.O. Caldeira and A.J. Leggett}
}

@article{feynman1963theory,
title = {The theory of a general quantum system interacting with a linear dissipative system},
author = {Feynman, R. P. and Vernon, F. L.},
year = {1963},
volume = {24},
pages = {118--173},
journal = {Ann. Phys. },
doi = {10.1016/0003-4916(63)90068-X}
}

@book{breuer2002theory,
  title={The Theory of Open Quantum Systems},
  author={Breuer, H.-P. and Petruccione, F.},
  publisher={Oxford University Press},
  year={2002}
}

@article{linblad1976generators,
title = {On the generators of quantum dynamical semigroups},
author ={Lindblad, G.},
year ={1976},
volume = {48},
pages = {119--130},
journal = {Commun. Math. Phys},
doi = {10.1007/BF01608499}
}

@article{wiercinski_role_2024,
	title        = {Role of polaron dressing in superradiant emission dynamics},
	author       = {Wiercinski, J. and Cygorek, M. and Gauger, E. M.},
	year         = 2024,
	month        = sep,
	journal      = {Phys. Rev. Res.},
	volume       = 6,
	pages        = {033231},
	doi          = {10.1103/PhysRevResearch.6.033231}
}

@article{lacroix2026tensornetworkmethodsnonperturbative,
      title={Tensor network methods for non-perturbative dynamics of open quantum systems}, 
      author={Thibaut Lacroix and Adam Burgess and Nicola Lorenzoni and Julian Wiercinski and Kian Damezin and James Lim and Dario Tamascelli and Alex W. Chin and Moritz Cygorek and Brendon W. Lovett and Jonathan Keeling and Susana F. Huelga and Martin B. Plenio and Erik M. Gauger},
      year={2026},
       journal       = {arXiv preprint},
      doi = {10.48550/arXiv.2608.09850}
}

@article{wang2003multilayer,
    author = {Wang, Haobin and Thoss, Michael},
    title = {Multilayer formulation of the multiconfiguration time-dependent Hartree theory},
    journal = {J. Chem. Phys.},
    volume = {119},
    number = {3},
    pages = {1289-1299},
    year = {2003},
    month = {07},
    issn = {0021-9606},
    doi = {10.1063/1.1580111},
   }

@article{shi2009efficient,
title ={Efficient hierarchical Liouville space propagator to quantum dissipative dynamics},
author = {Shi, Q. and Chen, L.P. and Nan, G. and Xu, R.X. and Yan, Y.J.},
year={2009},
volume={130},
pages={084105},
journal={J. Chem. Phys.},
doi = {10.1063/1.3077918}
}

@article{Wang2007,
  title={Multilayer formulation of the multiconfiguration time-dependent Hartree theory},
  author={Wang, H. and Thoss, M.},
  year={2003},
  volume={119},
  pages={1289--1299},
  journal={J. Chem. Phys.},
  doi = {10.1063/1.1580111}
}

@article{MaxInesDeVega,
    title = {Dephasing dynamics of an impurity coupled to an anharmonic environment},
    author = {Bramberger, M and de Vega, I}, 
    year = {2020},
    volume = {101},
    pages = {012101},
    journal = {Phys. Rev. A},
    doi = {10.1103/PhysRevA.101.012101}}

@article{deVega2017,
    title={Dynamics of non-Markovian open quantum systems},
    author={de Vega, I. and Alonso, D.},
    year={2017},
    volume={89},
    pages={015001},
    journal={Rev. Mod. Phys.},
    doi={10.1103/RevModPhys.89.015001}
}

@article{CarlosDeVega,
    title={Open quantum systems in thermal nonergodic environments},
    author={Parra-Murillo, C. A. and Bramberger, M. and Hubig, C. and de Vega, I.},
    year={2021},
    volume={103},
    pages={032204},
    journal={Phys. Rev. A},
    doi={10.1103/PhysRevA.103.032204}
}

@article{Makri2024EssentialBathAnharmonicity,
    title={Essential vs Removable Bath Anharmonicity: Path Integral Results with Model Electronic-Vibrational Baths},
    author={Makri, N.},
    year={2024},
    volume={15},
    pages={9766--9773},
    journal={J. Phys. Chem. Lett.},
    doi={10.1021/acs.jpclett.4c02213}
}

@article{Ye2021TensorNetworkIF,
    title={Constructing Tensor Network Influence Functionals for General Quantum Dynamics},
    author={Ye, E. and Chan, G. K.-L.},
    year={2021},
    volume={155},
    pages={044104},
    journal={J. Chem. Phys.},
    doi={10.1063/5.0047260}
}

@article{Smith2019QuantumDissipative,
    title={Quantum dissipative systems beyond the standard harmonic model: Features of linear absorption and dynamics},
    author={Smith, L. D. and Dijkstra, A. G.},
    year={2019},
    volume={151},
    pages={164109},
    journal={J. Chem. Phys.},
    doi={10.1063/1.5122896}
}

@article{tiwari2013electronic,
    title={Electronic resonance with anticorrelated pigment vibrations drives photosynthetic energy transfer outside the adiabatic framework},
    author={Tiwari, V. and Peters, W. K. and Jonas, D. M.},
    year={2013},
    volume={110},
    pages={1203--1208},
    journal={PNAS},
    doi = {10.1073/pnas.1211157110}
}

@book{mukamel1995nonlinear,
    title={Principles of Nonlinear Optical Spectroscopy},
    author={Mukamel, S.},
    year={1995},
    publisher={Oxford University Press}
}

@article{scholes2003long,
    title={Long-range resonance energy transfer in molecular systems},
    author={Scholes, G. D.},
    year={2003},
    volume={54},
    pages={57--87},
    journal={Annu. Rev. Phys. Chem.},
    doi = {10.1146/annurev.physchem.54.011002.103746}
}

@article{cho2008coherent,
    title={Coherent two-dimensional optical spectroscopy},
    author={Cho, M.},
    year={2008},
    volume={108},
    pages={1331--1418},
    journal={Chem. Rev.},
    doi = {10.1021/cr078377b}
}

@article{makri1999linear,
author = {Makri, Nancy},
title = {The Linear Response Approximation and Its Lowest Order Corrections: An Influence Functional Approach},
journal = { J. Phys. Chem. B},
volume = {103},
number = {15},
pages = {2823-2829},
year = {1999},
doi = {10.1021/jp9847540},


}

@article{delima2005morse,
  author = {E. F. de Lima and J. E. M. Hornos},
  title = {Matrix elements for the Morse potential under an external field},
  journal = {J. Phys. B: At. Mol. Opt. Phys.},
  volume = {38},
  pages = {815--822},
  year = {2005},
  doi = {10.1088/0953-4075/38/7/004}
}

@article{ENMiranda_2001,
doi = {10.1088/0143-0807/22/5/303},
url = {10.1088/0143-0807/22/5/303},
year = {2001},
month = {jul},
publisher = {},
volume = {22},
number = {5},
pages = {483},
author = {E N Miranda},
title = {A paradox in the electronic partition function or how to be cautious
with mathematics},
journal = {Eur. J. Phys.},
}

@article{mortiz2022ACE,
    title = {Simulation of open quantum systems by automated compression of arbitrary environments},
    author = {Cygorek, M and Cosacchi, M and Vagov, A and Axt, M, V and Lovett, W, B and Keeling, J and Gauger, M, E}, 
    journal = {Nat. Phys.},
    volume = {18},
    pages = {662 - 668},
    year = {2022},
    doi = {10.1038/s41567-022-01544-9}
}

@article{DAMPF,
  title = {Dissipation-Assisted Matrix Product Factorization},
  author = {Somoza, Alejandro D. and Marty, Oliver and Lim, James and Huelga, Susana F. and Plenio, Martin B.},
  journal = {Phys. Rev. Lett.},
  volume = {123},
  issue = {10},
  pages = {100502},
  numpages = {7},
  year = {2019},
  month = {Sep},
  publisher = {American Physical Society},
  doi = {10.1103/PhysRevLett.123.100502},
 
}

@article{NicolaSystematicCoarseGraining,
  title = {Systematic Coarse Graining of Environments for the Nonperturbative Simulation of Open Quantum Systems},
  author = {Lorenzoni, Nicola and Cho, Namgee and Lim, James and Tamascelli, Dario and Huelga, Susana F. and Plenio, Martin B.},
  journal = {Phys. Rev. Lett.},
  volume = {132},
  issue = {10},
  pages = {100403},
  numpages = {7},
  year = {2024},
  month = {Mar},
  publisher = {American Physical Society},
  doi = {10.1103/PhysRevLett.132.100403}
}

@article{FEYNMAN1963118,
	title        = {The theory of a general quantum system interacting with a linear dissipative system},
	author       = {R.P Feynman and F.L Vernon},
	year         = 1963,
	journal      = {Ann. Phys. },
	volume       = 24,
	pages        = {118--173},
	doi          = {10.1016/0003-4916(63)90068-X},
	issn         = {0003-4916},
	url          = {https://www.sciencedirect.com/science/article/pii/000349166390068X}
}

@article{jorgensen_exploiting_2019,
	title        = {Exploiting the {Causal} {Tensor} {Network} {Structure} of {Quantum} {Processes} to {Efficiently} {Simulate} {Non}-{Markovian} {Path} {Integrals}},
	author       = {Jorgensen, Mathias R. and Pollock, Felix A.},
	year         = 2019,
	journal      = {Phys. Rev. Lett.},
	volume       = 123,
	number       = 24,
	pages        = 240602,
	doi          = {10.1103/PhysRevLett.123.240602},
	url          = {https://link.aps.org/doi/10.1103/PhysRevLett.123.240602}
}

@article{tamascelli_efficient_2019,
	title        = {Efficient {Simulation} of {Finite}-{Temperature} {Open} {Quantum} {Systems}},
	author       = {Tamascelli, D. and Smirne, A. and Lim, J. and Huelga, S. F. and Plenio, M. B.},
	year         = 2019,
	month        = aug,
	journal      = {Phys. Rev. Lett.},
	volume       = 123,
	number       = 9,
	pages        = {090402},
	doi          = {10.1103/PhysRevLett.123.090402},
	issn         = {0031-9007, 1079-7114},

}

@article{Lopez2011,
    author = {López-López, S. and Martinazzo, R. and Nest, M.},
    title = {Benchmark calculations for dissipative dynamics of a system coupled to an anharmonic bath with the multiconfiguration time-dependent Hartree method},
    journal = {J. Chem. Phys.},
    volume = {134},
    number = {9},
    pages = {094102},
    year = {2011},
    month = {03},
    issn = {0021-9606},
    doi = {10.1063/1.3556940},
    url = {10.1063/1.3556940},

}

@article{Scholes2011,
  author = {Scholes, Gregory D. and Fleming, Graham R. and Olaya‑Castro, A. and van Grondelle, R.},
  title = {Lessons from nature about solar light harvesting},
  journal = {Nat. Chem.},
  volume = {3},
  pages = {763–774},
  year = {2011},
  doi = {10.1038/nchem.1145}
}

@article{Yeganeh2006,
  author = {Yeganeh, Sina and Ratner, Mark A.},
  title = {Effects of anharmonicity on nonadiabatic electron transfer: A model},
  journal = {J. Chem. Phys.},
  volume = {124},
  number = {4},
  pages = {044108},
  year = {2006},
  doi = {10.1063/1.2162172},
  url = {10.1063/1.2162172}
}

@article{Evans2000,
  author = {Evans, Deborah G.},
  title = {Anharmonic effects in photoinduced electron transfer},
  journal = {J. Chem. Phys.},
  volume = {113},
  pages = {3282--3288},
  year = {2000},
  doi = {10.1063/1.1286961},
  url = {10.1063/1.1286961}
}

@article{WangThoss2004,
  author = {Wang, Haobin and Thoss, Michael},
  title = {Semiclassical simulation of absorption spectra for a chromophore coupled to an anharmonic bath},
  journal = {Chem. Phys.},
  volume = {304},
  number = {1-2},
  pages = {121--131},
  year = {2004},
  doi = {10.1016/j.chemphys.2004.06.007},
  url = {10.1016/j.chemphys.2004.06.007}
}

@article{ThossWang2006,
  author = {Thoss, Michael and Wang, Haobin},
  title = {Quantum dynamical simulation of ultrafast molecular processes in the condensed phase},
  journal = {Chem. Phys.},
  volume = {322},
  number = {1-2},
  pages = {210--222},
  year = {2006},
  doi = {10.1016/j.chemphys.2005.07.011},
  url = {10.1016/j.chemphys.2005.07.011}
}

@article{Jonas2003,
  author    = {Jonas, David M.},
  title     = {Two-dimensional femtosecond spectroscopy},
  journal   = {Annu. Rev. Phys. Chem.},
  volume    = {54},
  pages     = {425--463},
  year      = {2003},
  doi       = {10.1146/annurev.physchem.54.011002.103907}
}

@article{Engel2007,
  author    = {Engel, Gregory S. and Calhoun, Tessa R. and Read, Elizabeth L. and Ahn, Tae-Kyu and Man\v{c}al, Tom\'{a}\v{s} and Cheng, Yuan-Chung and Blankenship, Robert E. and Fleming, Graham R.},
  title     = {Evidence for wavelike energy transfer through quantum coherence in photosynthetic systems},
  journal   = {Nature},
  volume    = {446},
  pages     = {782--786},
  year      = {2007},
  doi       = {10.1038/nature05678}
}

@article{BurgessGauger2026,
  author       = {Adam Burgess and Erik M. Gauger},
  title        = {Enhancing energy transport utilising permanent molecular dipoles},
  journal      = {Phys. Chem. Chem. Phys.},
  year         = {2026},
  volume       = {28},
  pages        = {1792--1805},
  doi          = {10.1039/D5CP04076K},
  note         = {Advance Article, first published 17 December 2025}
}

@article{AChin_ENAQT,
  title = {Efficient Biologically Inspired Photocell Enhanced by Delocalized Quantum States},
  author = {Creatore, C. and Parker, M. A. and Emmott, S. and Chin, A. W.},
  journal = {Phys. Rev. Lett.},
  volume = {111},
  issue = {25},
  pages = {253601},
  numpages = {5},
  year = {2013},
  month = {Dec},
  publisher = {American Physical Society},
  doi = {10.1103/PhysRevLett.111.253601},
  url = {https://link.aps.org/doi/10.1103/PhysRevLett.111.253601}
}

@article{RouseGaugerLovett2019,
  author       = {D.~M.~Rouse and E.~M.~Gauger and B.~W.~Lovett},
  title        = {Optimal power generation using dark states in dimers strongly coupled to their environment},
  journal      = {New J. Phys.},
  year         = {2019},
  volume       = {21},
  number       = {6},
  pages        = {063025},
  doi          = {10.1088/1367-2630/ab25ca}
}

@article{PlenioHuelga2008,
  author       = {M. B. Plenio and S. F. Huelga},
  title        = {Dephasing‐assisted transport: quantum networks and biomolecules},
  journal      = {New J. Phys.},
  volume       = {10},
  year         = {2008},
  pages        = {113019},
  doi          = {10.1088/1367-2630/10/11/113019}
}

@article{ENAQT,
  author       = {Patrick Rebentrost and Masoud Mohseni and Ivan Kassal and Seth Lloyd and Al{\'a}n Aspuru‐Guzik},
  title        = {Environment‐Assisted Quantum Transport},
  journal      = {New J. Phys.},
  year         = {2009},
  volume       = {11},
  pages        = {033003},
  doi          = {10.1088/1367-2630/11/3/033003}
}

@article{Zerah-HarushDubi2018,
  author       = {Elinor Zerah-Harush and Yonatan Dubi},
  title        = {Universal Origin for Environment‐Assisted Quantum Transport in Exciton Transfer Networks},
  journal      = {J. Phys. Chem. Lett.},
  year         = {2018},
  volume       = {9},
  number       = {7},
  pages        = {1689--1695},
  doi          = {10.1021/acs.jpclett.7b03306}
}

@article{Sowa2017,
  author = {J. K. Sowa and J. A. Mol and G. A. D. Briggs and E. M. Gauger},
  title = {Environment-assisted quantum transport through single-molecule junctions},
  journal = {Phys. Chem. Chem. Phys.},
  year = {2017},
  volume = {19},
  pages = {29534--29539},
  doi = {10.1039/C7CP06237K}
}

@article{Biggerstaff2016,
  author = {D. N. Biggerstaff and R. Heilmann and A. A. Zecevik and M. Grafe and M. A. Broome and A. Fedrizzi and S. Nolte and A. Szameit and A. G. White and I. Kassal},
  title = {Enhancing coherent transport in a photonic network using controllable decoherence},
  journal = {Nat. Commun.},
  year = {2016},
  volume = {7},
  pages = {11282},
  doi = {10.1038/ncomms11282}
}

@article{chin_exact_2010,
	title        = {Exact mapping between system-reservoir quantum models and semi-infinite discrete chains using orthogonal polynomials},
	author       = {Chin, Alex W. and Rivas, Angel and Huelga, Susana F. and Plenio, Martin B.},
	year         = 2010,
	month        = sep,
	journal      = {J. Math. Phys.},
	volume       = 51,
	number       = 9,
	pages        = {092109},
	doi          = {10.1063/1.3490188},
	issn         = {0022-2488, 1089-7658},
}

@article{strathearn_efficient_2018,
	title        = {Efficient non-{Markovian} quantum dynamics using time-evolving matrix product operators},
	author       = {Strathearn, A. and Kirton, P. and Kilda, D. and Keeling, J. and Lovett, B. W.},
	year         = 2018,
	month        = dec,
	journal      = {Nat. Commun.},
	volume       = 9,
	number       = 1,
	pages        = 3322,
	doi          = {10.1038/s41467-018-05617-3},
	issn         = {2041-1723},
	url          = {http://www.nature.com/articles/s41467-018-05617-3}
}

@article{Kubo1962,
author = {Kubo ,Ryogo},
title = {Generalized Cumulant Expansion Method},
journal = {J. Phys. Soc. Jpn.},
volume = {17},
number = {7},
pages = {1100-1120},
year = {1962},
doi = {10.1143/JPSJ.17.1100}
}

@article{Dubertrand_solid_state,
  title = {Analytical results for the quantum non-Markovianity of spin ensembles undergoing pure dephasing dynamics},
  author = {Dubertrand, R\'emy and Cesa, Alexandre and Martin, John},
  journal = {Phys. Rev. A},
  volume = {97},
  issue = {6},
  pages = {062126},
  numpages = {10},
  year = {2018},
  month = {Jun},
  publisher = {American Physical Society},
  doi = {10.1103/PhysRevA.97.062126},
  url = {https://link.aps.org/doi/10.1103/PhysRevA.97.062126}
}

@article{chekhovich_nuclear_2013,
	title = {Nuclear spin effects in semiconductor quantum dots},
	volume = {12},
	issn = {1476-4660},
	url = {10.1038/nmat3652},
	doi = {10.1038/nmat3652},
	number = {6},
	journal = {Nat. Mater.},
	author = {Chekhovich, E. A. and Makhonin, M. N. and Tartakovskii, A. I. and Yacoby, A. and Bluhm, H. and Nowack, K. C. and Vandersypen, L. M. K.},
	month = jun,
	year = {2013},
	pages = {494--504},
}

@article{LacorixPhotonicStructure,
  doi = {10.22331/q-2024-04-03-1305},
  title = {From {N}on-{M}arkovian {D}issipation to {S}patiotemporal {C}ontrol of {Q}uantum {N}anodevices},
  author = {Lacroix, Thibaut and Lovett, Brendon W. and Chin, Alex W.},
  journal = {{Quantum}},
  issn = {2521-327X},
  publisher = {{Verein zur F{\"{o}}rderung des Open Access Publizierens in den Quantenwissenschaften}},
  volume = {8},
  pages = {1305},
  month = apr,
  year = {2024}
}

@article{schroder_tensor_2019,
	title = {Tensor network simulation of multi-environmental open quantum dynamics via machine learning and entanglement renormalisation},
	volume = {10},
	issn = {2041-1723},
	url = {10.1038/s41467-019-09039-7},
	doi = {10.1038/s41467-019-09039-7},
	number = {1},
	journal = {Nat. Commun.},
	author = {Schröder, Florian A. Y. N. and Turban, David H. P. and Musser, Andrew J. and Hine, Nicholas D. M. and Chin, Alex W.},
	month = mar,
	year = {2019},
	pages = {1062},
}

@article{NicolaMicroscopicSimulations,
author = {Nicola Lorenzoni  and Thibaut Lacroix  and James Lim  and Dario Tamascelli  and Susana F. Huelga  and Martin B. Plenio },
title = {Full microscopic simulations uncover persistent quantum effects in primary photosynthesis},
journal = {Sci. Adv.},
volume = {11},
number = {40},
pages = {eady6751},
year = {2025},
doi = {10.1126/sciadv.ady6751}}

@article{Cresser01111992,
author = {J.D. Cresser},
title = {Thermal Equilibrium in the Jaynes-Cummings Model},
journal = {J. Mod. Opt.},
volume = {39},
number = {11},
pages = {2187--2192},
year = {1992},
publisher = {Taylor \& Francis},
doi = {10.1080/09500349214552211},

}

@article{HOLSTEIN1959325,
title = {Studies of polaron motion: Part I. The molecular-crystal model},
journal = {Ann. Phys.},
volume = {8},
number = {3},
pages = {325-342},
year = {1959},
issn = {0003-4916},
doi = {10.1016/0003-4916(59)90002-8},

author = {T Holstein}
}

@article{kubo_generalized_1962,
	title        = {Generalized Cumulant Expansion Method},
	author       = {Kubo, Ryogo},
	year         = 1962,
	month        = jul,
	journal      = {J. Phys. Soc. Jpn.},
	volume       = 17,
	number       = 7,
	pages        = {1100--1120},
	doi          = {10.1143/JPSJ.17.1100},
	issn         = {0031-9015, 1347-4073},
}

@book{milonni_introduction_2019,
	title        = {An Introduction to Quantum Optics and Quantum Fluctuations},
	author       = {Milonni, Peter W.},
	year         = 2019,
	publisher    = {Oxford University Press},
	doi          = {10.1093/oso/9780199215614.001.0001},
	isbn         = {9780199215614},
}

@article{funo_dynamics_2024,
	title        = {Dynamics of a Quantum System Interacting with White Non-Gaussian Baths: Poisson Noise Master Equation},
	author       = {Funo, Ken and Ishizaki, Akihito},
	year         = 2024,
	month        = apr,
	journal      = {Phys. Rev. Lett.},
	volume       = 132,
	number       = 17,
	pages        = {170402},
	doi          = {10.1103/PhysRevLett.132.170402},
}

@article{feynman_theory_1963,
	title        = {The theory of a general quantum system interacting with a linear dissipative system},
	author       = {Feynman, R. P. and Vernon, F. L.},
	year         = 1963,
	journal      = {Ann. Phys. (N.Y.)},
	volume       = 24,
	pages        = {118--173},
	doi          = {10.1016/0003-4916(63)90068-X}
}

@article{whitenongaussianbath,
  title = {Dynamics of a Quantum System Interacting with White Non-Gaussian Baths: Poisson Noise Master Equation},
  author = {Funo, Ken and Ishizaki, Akihito},
  journal = {Phys. Rev. Lett.},
  volume = {132},
  issue = {17},
  pages = {170402},
  numpages = {7},
  year = {2024},
  month = {Apr},
  publisher = {American Physical Society},
  doi = {10.1103/PhysRevLett.132.170402}
}

@Article{Kramer2015,
author={Kr{\"a}mer, Sebastian
and Ritsch, Helmut},
title={Generalized mean-field approach to simulate the dynamics of large open spin ensembles with long range interactions},
journal={Eur. Phys. J. D.},
year={2015},
month={Dec},
day={17},
volume={69},
number={12},
pages={282},
issn={1434-6079},
doi={10.1140/epjd/e2015-60266-5}

}

@article{Roos,
  title = {Process tensor approaches to modeling two-dimensional spectroscopy},
  author = {de Wit, Roosmarijn and Keeling, Jonathan and Lovett, Brendon W. and Chin, Alex W.},
  journal = {Phys. Rev. Res.},
  volume = {7},
  issue = {1},
  pages = {013209},
  numpages = {10},
  year = {2025},
  month = {Feb},
  publisher = {American Physical Society},
  doi = {10.1103/PhysRevResearch.7.013209},

}

@article{
lorenzoni_persistent_2025,
author = {Nicola Lorenzoni  and Thibaut Lacroix  and James Lim  and Dario Tamascelli  and Susana F. Huelga  and Martin B. Plenio },
title = {Full microscopic simulations uncover persistent quantum effects in primary photosynthesis},
journal = {Sci. Adv.},
volume = {11},
number = {40},
pages = {eady6751},
year = {2025},
doi = {10.1126/sciadv.ady6751}}

@article{dewit_extracting_2015,
	title        = {Extracting coupling-mode spectral densities with two-dimensional electronic spectroscopy},
	author       = {de Wit, Roosmarijn and Keeling, Jonathan and Lovett, Brendon W. and Chin, Alex W.},
	year         = 2015,
	journal      = {J. Chem. Phys.},
	volume       = 143,
	number       = 21,
	pages        = 214107,
	doi          = {10.1063/1.4936884},

}

@article{InesDeVega_Daniel_Alonso_Dynmaics_OQS,
  title = {Dynamics of non-Markovian open quantum systems},
  author = {de Vega, In\'es and Alonso, Daniel},
  journal = {Rev. Mod. Phys.},
  volume = {89},
  issue = {1},
  pages = {015001},
  numpages = {58},
  year = {2017},
  month = {Jan},
  publisher = {American Physical Society},
  doi = {10.1103/RevModPhys.89.015001},
}
\end{document}